\documentclass{ws-ijmpa}
\usepackage {graphicx} 
\usepackage{amsmath}
\begin{document}

\markboth{Zaninetti}
{Cosmic Rays from Super-bubbles}

%
\catchline{}{}{}{}{}
%
\def\aplett{Astrophys.~Lett.\,}
\def\apj{ApJ\,}
\def\apjl{ApJ\,}
\def\aap{A\&A\,}
\def\mnras{MNRAS\,}
\def\pasj{PASJ\,}
\def\solphys{Sol.~Phys.\,}
\def\aj{AJ}
\def\aaps{A\&AS}
\def\araa{ARA\&A}
\def\apjs{ApJS}
\def\cjaa{Chinese J. Astron. Astrophys.}
\def\jaa{J. Astrophys. Astr.}
\def\planss{Planet.~Space~Sci.}
\def\apss{Astrophysics and Space Science}
\def\pasp{PASP}
\def\JPG{J. Phys. G\,}
\def\POF{Physics of Fluids}
\def\physrep{Phys.~Rep.\,}
\def\nat{Nature\,}
\def\pre{Phys.~Rev.~E}
\def\pra{Phys.~Rev.~A}
\def\physa{Physica A}
\def\rpp{Rep.Prog.Phys.}
\title
{Models of diffusion of Galactic Cosmic
                Rays from Super-bubbles }

\author{ Lorenzo Zaninetti}

\address {Dipartimento di Fisica Generale, Via Pietro Giuria 1,\\
           I-10125 Torino, Italy}

\maketitle

\begin{history}
\received{Day Month Year}
\revised{Day Month Year}
\end{history}

\begin{abstract}

Super-bubbles are shells in the interstellar medium produced
by the simultaneous explosions of many supernova remnants. 
The solutions of the mathematical diffusion 
and of the Fourier expansion in 1D, 2D and 3D were
deduced in
order to describe the diffusion of nucleons from such structures. 
The mean number of visits in the the case of the Levy flights in 1D
was computed  with a Monte Carlo simulation.
The diffusion of cosmic rays has its  physical explanation
in the relativistic Larmor gyro-radius  which  
is energy dependent.
The mathematical solution of the diffusion equation in 1D with
variable diffusion coefficient was computed. Variable diffusion
coefficient means  magnetic field  variable with the altitude from
the Galactic plane. The analytical solutions allow  us  to calibrate
the code that describes the Monte Carlo diffusion.
 The maximum energy that can be 
extracted from the super-bubbles is  deduced. The concentration of
cosmic rays is a function of the distance from the nearest
super-bubble and the selected
energy. 
The
interaction of the cosmic rays on the target material allows  us  to
trace the theoretical map of the diffuse Galactic continuum
gamma-rays.
The streaming of the cosmic rays from the Gould Belts that contains
the sun at it's internal was described by a Monte Carlo simulation.
Ten new formulas are derived. 

\end{abstract}

\ccode{PACS numbers: 96.40.-z , 96.50.Pw , 05.40.Fb}

\section{Introduction}

The spatial diffusion of  cosmic rays (CR) plays a relevant
role in astrophysics due to the fact that the binary collisions
between CR and interstellar medium are negligible, e.g.
\cite{Kulsrud2005,Sokolosky2004}. In these last years a
general consensus has been reached on the fact that particle
acceleration in Supernova Remnants , in the following SNR,
  may provide   CR at energies up to
$10^{15}eV$ \cite {Blandford1987} and probability density function
in energy p(E)$\propto E^{-2}$. Some of  methods
for the numerical
computation of the propagation of primary and secondary nucleons
have  been implemented \cite{Strong1998,Taillet2003}~. 
The
previous efforts have covered the diffusion of CR from SNR
\cite{Lingenfelter1969,Schlickeiser2005,Wolfendale_2006}  but the SNR are 
often concentrated in
super-bubbles (SB) that may reach up to 800~pc in altitude from
the Galactic plane.  The complex 3D structure of the SB represents
the spatial coordinates where the CR are injected. 
We remember that the 
super-bubbles are the likely source of at least a
substantial fraction of GCRs; 
this can be   shown  from the analysis of  the  data 
from the Cosmic Ray Isotope Spectrometer (CRIS) aboard the
Advanced Composition Explorer (ACE) 
and in particular   the 22Ne/20Ne ratio at the cosmic-ray 
source using the
measured 21Ne, 19F, and 17O abundances as `tracers'' 
of secondary production \cite{Binns2005}.
Further on  is  known  that  $\approx~ 75\%$ of supernova occur in
super-bubbles, and   $\approx~ 88 \%$  
of the cosmic-ray heavy particles are accelerated there
because of the factor of ~3 enhanced super-bubble core metallicity
\cite{Higdon2005,Higdon2003}~.

In order to
describe the propagation of CR from SB,  in
Section~\ref{sec_mat} we have developed 
 the mathematical and Fourier solutions of the diffusion
in presence of the stationary state and the
 Monte Carlo simulations   which  allow us to solve the equation of the diffusion 
from a numerical
 point of view .
 The case of random walk with variable length of the step
 was analysed in Section~\ref{levysec}.
 The physical nature of the diffusion as well as
 evaluations on transit times and spectral index
 are discussed in Section~\ref{physics}.
 The 3D diffusion with a fixed length of a step and injection points
 randomly chosen on the surface of the
 expanding  SB was treated in Section~\ref{sb}.

\section{Preliminaries}
\label{sec_mat}
In this section the  basic equations of  transport are summarised,
the rules that govern the Monte Carlo diffusion are set up
and the various solutions of the mathematical and Fourier diffusion
are introduced.
The statistics of
the visits to the sites in the theory
of Levy flights and it's connection with the regular random walk 
are analysed.
It is  important to point out that the mathematical and Monte Carlo
diffusion here considered cover the stationary situation.

\subsection{Particle transport equations}
The diffusion-loss  equation as deduced by
\cite{longair} (~equation~(20.1)~)
for light nuclei  once the spallation phenomena
are neglected  has the form
\begin {equation}
\frac  {\partial   N_i } {\partial  t} =
D \nabla ^2 N_i   +
\frac {\partial } {\partial E}  [ b(E)N_i]
+ Q_i
 ,
\label{fundamental}
\end {equation}
here $N_i$ is the number density of nuclei of species i ,
$\nabla ^2$ is the Laplacian operator ,
D is the scalar diffusion coefficient,
$\frac {\partial}{\partial E}  [ b(E)N_i]$
takes account of the energy balance,
and
$Q_i$ represents  the  injection rate  per
unit volume.

 When only  protons
are considered and  the energy dependence is neglected
Equation~(\ref{fundamental}) becomes
\begin {equation}
\frac  {\partial P(x,y,z,t)} {\partial t} =
 D \nabla^2  P (x,y,z,t) \quad ,
\label{diffusion3d}
\end {equation}
where  P is the probability density~,
see~\cite{gould}
(equation~12.34b~).
In this  formulation the
dependence for the mean square displacement $\overline{R^2(t)}$
is, see~\cite{gould} (equation.~8.38~),:
\begin {equation}
\overline {R^2(t)} = 2d Dt \quad (t \rightarrow \infty ) \quad ,
\label{r2_continuum}
\end   {equation}
where d  is the spatial dimension  considered.
From equation~(\ref{r2_continuum}), the  diffusion coefficient
is derived
in the continuum :
\begin {equation}
D  =  (t \rightarrow \infty) \frac {\overline R^2}{2dt}
\label{Dt}
\quad  .
\end   {equation}
In the presence of discrete time steps on a 3D lattice the average
square radius after N steps,
 see~\cite {gould}
(equation~12.5~), is
\begin {equation}
\langle R^2(N  )\rangle  \sim 2d  DN \quad ,
\label{rquadro}
\end   {equation}
from which the diffusion coefficient is derived
\begin {equation}
D= \frac {\langle R^2(N  )\rangle} {  2d N} \quad .
\label{r2N}
\end   {equation}
If ${\langle R^2(N  )\rangle}~\sim~N$ , as  in our lattice :
\begin {equation}
D= \frac {1} {  2d  } \quad ,
\end   {equation}
when the physical units are 1 and
\begin {equation}
D= \frac {1} {  2d }  \lambda v_{tr}   \quad ,
\label{coefficient}
\end   {equation}
when the step   length of the walker is $\lambda$ and the
transport velocity  $v_{tr}$.

Another useful  law is   Fick'~s second equation
in three dimensions~\cite{berg},
\begin{equation}
\frac {\partial C }{\partial t} =
D \nabla^2 C
\quad ,
\label{eqfick}
\end {equation}
where the concentration C(x,y,z) is  the number of particles per
unit volume at the point (x,y,z).
We now summarise how Fick'~s second equation
transforms itself when the steady state is considered
in  the light of the three fundamental symmetries.
In the presence of  spherical symmetry,
  equation~(\ref{eqfick})
becomes
\begin{equation}
\frac{1}{r^2}
\frac{d}{dr}
( r^2 \frac {dC}{dr})
=0
\quad ,
\label{fick_3D}
\end {equation}
the  circular  symmetry   (starting from the hollow cylinder)
gives
\begin{equation}
\frac{d}{dr}
( r \frac {dC}{dr})
=0
\quad ,
\label{fick_2D}
\end {equation}
and the  point  symmetry   (diffusion from a plane)
produces
\begin{equation}
\frac {d^2C}{dr^2} =0
\quad ,
\label{fick_1D}
\end {equation}
in the case of constant D and
\begin{equation}
\frac {d}{dr}(D\frac{dC}{dr}) =0
\quad ,
\label{fick_1D_var}
\end {equation}
when  D is a function of the distance or the concentration.


\subsection{Monte Carlo diffusion}

The adopted rules of the d-dimensional  (with d=1,2,3)
random walk on a  lattice characterised
by $NDIM^d$  elements occupying  a
physical volume $side^d$  (in the hypothesis of a 
stationary state and fixed length step)
are specified as follows:
\begin  {enumerate}
\item    The motion starts at the center of the lattice ,
\item    The particle   reaches one of the     $2 \times d $
         surroundings  and the  procedure repeats itself ,
\item    The motion terminates when one of the $2 \times d $
         boundaries   is reached ,
\item    The number of visits
         is   recorded on ${\mathcal M}^d$ , a d--dimensional
         memory or concentration  grid ,
\item    This  procedure is repeated  NTRIALS times
         starting from (1) with a different pattern ,
\item    For the sake  of normalisation the
         d--dimensional memory or concentration  grid
         ${\mathcal M}^d$ is divided by NTRIALS .
\end   {enumerate}

\subsection{Mathematical diffusion}

The solutions of the mathematical diffusion
are now derived
in  3D,2D,1D and 1D with variable 
diffusion coefficient.

\subsubsection{3D analytical solution}
\label{mathematical}
\begin{figure}[pb]
\includegraphics[width=12cm,angle=0]{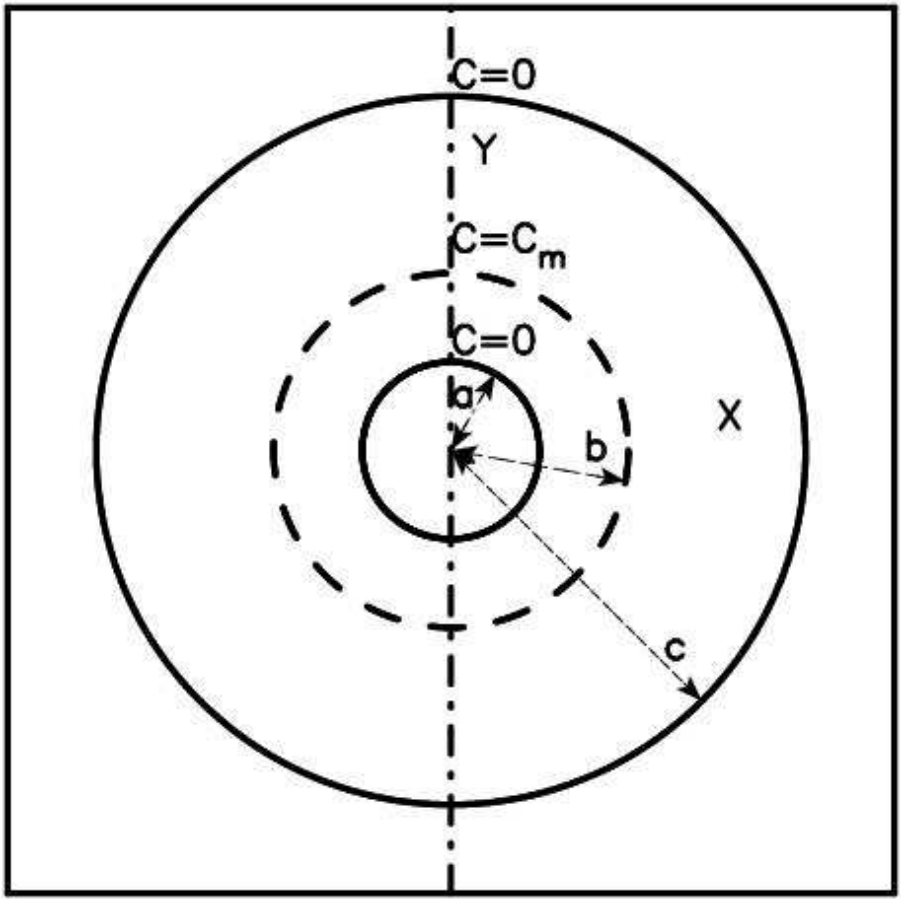}
\vspace*{8pt}
\caption {The spherical source is  represented  by 
a dashed line, the two absorbing boundaries
with a full line.
\label{plot}}
 \end{figure}

Consider a spherical shell  source of radius  b
between a spherical absorber
of radius a and a spherical absorber of radius c,
see Figure~\ref{plot} that is adapted from Figure~3.1 of~\cite{berg}~.

\begin{figure}
\includegraphics[width=12cm,angle=0]{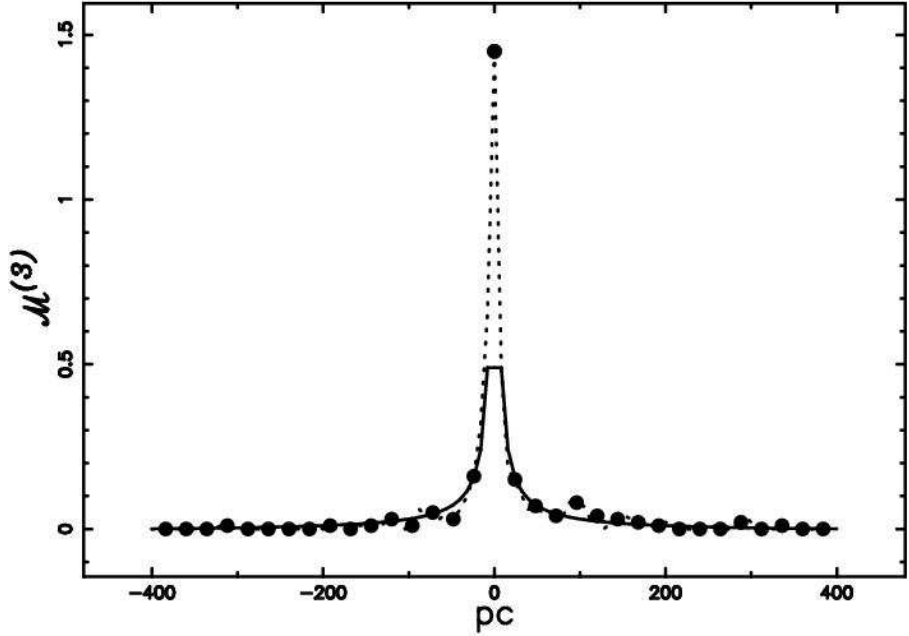}
\caption
{
Values of ${\mathcal M}^{(3)}$ or concentration  computed with
equation~(\ref{cbc}) (full line) compared with the results
of a Monte Carlo simulation
(filled circles).
The parameters are:
NDIM=101,
NTRIALS=100 and side=800pc.
\label{soluz_3d}}
    \end{figure}

The  concentration rises from 0 at r=a to a
maximum value $C_m$ at r=b and then  falls again
to 0 at r=c~.
The  solution of  equation~(\ref{fick_3D}) is
\begin{equation}
C(r) = A +\frac {B}{r}
\quad ,
\label{solution}
\end {equation}
where A and B  are determined from the boundary conditions~,
\begin{equation}
C(r) =
C_{{m}} \left( 1-{\frac {a}{r}} \right)  \left( 1-{\frac {a}{b}}
 \right) ^{-1}
\quad a \leq r \leq b
\quad ,
\end{equation}
and
\begin{equation}
C(r)=
C_{{m}} \left( {\frac {c}{r}}-1 \right)  \left( {\frac {c}{b}}-1
 \right) ^{-1}
\quad b \leq r \leq c
\quad .
\label{cbc}
\end{equation}
These solutions can be found in ~\cite{berg} or in~\cite{crank}~.
The  second   solution (equation~(\ref{cbc}))
can be applied to explain the 3D diffusion from
a central point once  the following connection
between continuum and discrete random walk  is made
\begin{eqnarray}
b & = & \frac {side}{NDIM-1}  \\ \nonumber
c & = & \frac {side}{NDIM-1} (\frac {NDIM+1}{2} -1 )
\quad  .
\end{eqnarray}
For example when $NDIM~\gg~1$ or $\frac{c}{b}~\gg~1$
\begin {equation}
C(r) \sim C_{{m}}\frac {b}{r}
\label{approximate3D}
\quad ,
\end {equation}
in other words, the concentration scales as an
hyperbola with  unity represented by the
mean free path b.
The 3D theoretical solution  as well as the Monte Carlo
simulation are reported in Figure~\ref{soluz_3d}.

\subsubsection{2D analytical solution}

\begin{figure}
\includegraphics[width=12cm,angle=0]{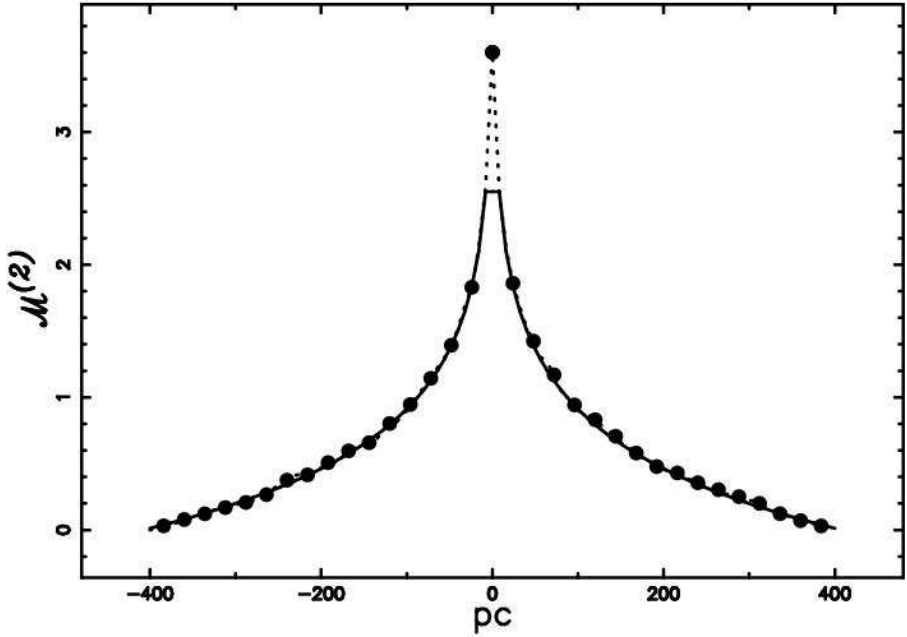}
\caption
{
Values of ${\mathcal M}^{(2)}$ or concentration  computed with
equation~(\ref{cbc_2d}) (full line) compared with the results
of a Monte Carlo simulation
(filled circles).
The parameters are:
NDIM=101,
NTRIALS=4000 and side=800pc.
\label{soluz_2d}}
    \end{figure}

The 2D solution is the same as 
the  hollow cylinder.
The general solution to   equation~(\ref{fick_2D})
is
\begin{equation}
C(r) = A + B ln(r)
\quad .
\label{solution_2D}
\end{equation}
The boundary conditions  give 
\begin{equation}
C(r) =
C_{{m}}  \frac {ln(r/a)} {ln(b/a)}
\quad a \leq r \leq b
\quad ,
\end{equation}
and
\begin{equation}
C(r) =
C_{{m}}  \frac {ln(r/c)} {ln(b/c)}
\quad b \leq r \leq c
\quad .
\label{cbc_2d}
\end{equation}
The new 2D theoretical solution  as well as the Monte Carlo
simulation are reported in Figure~\ref{soluz_2d}.

\subsubsection{1D analytical solution}

\begin{figure}
\includegraphics[width=12cm,angle=0]{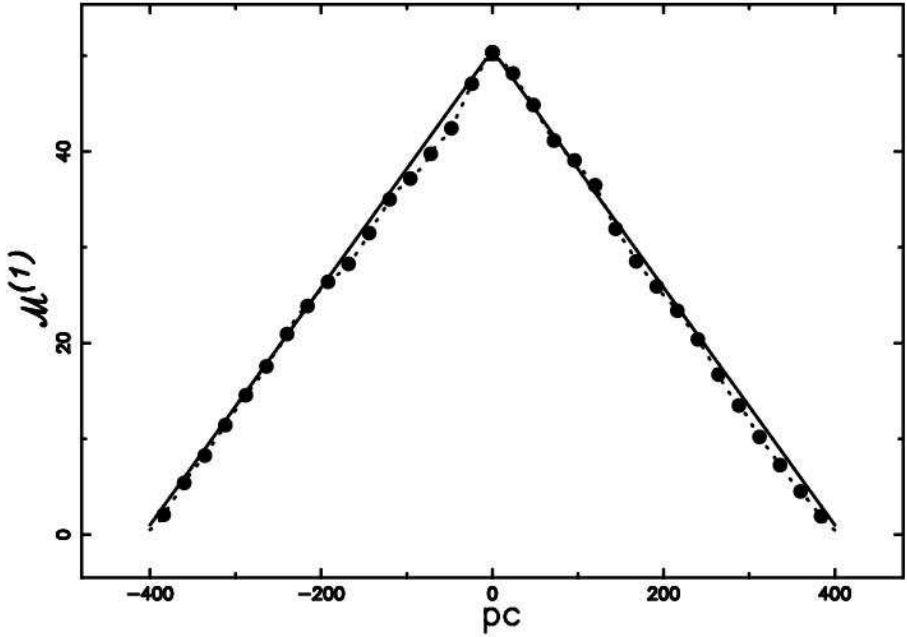}
\caption
{
Values of ${\mathcal M}^{(1)}$ or concentration  computed with
equation~(\ref{cbc_1d}) (full line) compared with the results
of a Monte Carlo simulation
(filled circles).
The parameters are:
NDIM=101,
NTRIALS=100 and side=800pc.
\label{soluz_1d}}
    \end{figure}

The 1D solution is the same as 
the diffusion through  a plane sheet.
The general solution to   equation~(\ref{fick_1D})
is
\begin{equation}
C(r) = A + B r
\quad .
\label{solution_1D}
\end{equation}
The boundary conditions  give
\begin{equation}
C(r) =
C_{{m}}  \frac {r-a}{b-a}
\quad a \leq r \leq b
\quad ,
\end{equation}
and
\begin{equation}
C(r) =
C_{{m}}   \frac {r-c}{b-c}
\quad b \leq r \leq c
\quad .
\label{cbc_1d}
\end{equation}
The 1D theoretical solution  as well as the Monte Carlo
simulation are reported in Figure~\ref{soluz_1d}.

\subsubsection{1D analytical solution with variable D}

\label{sec_diffvar}
The diffusion coefficient D is assumed to be a function ,f ,of
the distance
\begin{equation}
D=D_0 ( 1 + f(r))
\quad  ,
\end{equation}
where $D_0$ is the value of D at the plane  where the
diffusion starts.
On introducing the integral I defined as :
\begin{equation}
I =\int \frac  {dr}{1+f(r)}
\quad  ,
\end{equation}
the solution  to equation~(\ref{fick_1D_var})  is found to be  ,
\begin{equation}
\frac{C -C_1}{C_1 -C_2} =  \frac{I_1-I}{I_1-I2}
\quad ,
\end{equation}
where $C_1$ ,$C_2$ , $I_1$ and $I_2$ are the
 concentration and the integral  respectively
computed at $r=r_1$ and $r=r_2$   \cite{crank,barrer}.
For the sake of simplicity the diffusion coefficient is assumed
to be
\begin{equation}
D=D_0 (1+ a_D (r-r_1))
\quad ,
\label{D_variable}
\end{equation}
where $a_D$ is a coefficient that will be fixed in Section~\ref{dvariable}~.
A practical  way to determine $a_D$ is connected
with the interval of existence of r , $[r_1,r_2]$, and with the
value that D  takes at the boundary
\begin{equation}
a_D = \frac {D(r_2) /D_0 -1}{r_2 -r_1}
\quad.
\end{equation}
Along a line perpendicular
to the plane the  concentration rises from 0 at r=a to a
maximum value $C_m$ at r=b and then  falls again
to 0 at r=c~.
The variable diffusion coefficient  is
\begin{eqnarray}
D=D_0 (1 + a_D (b-r)) \\
when \quad a \leq r \leq b  \nonumber
\quad ,
\end{eqnarray}
and
\begin{eqnarray}
D=D_0 (1 + a_D (r-b)) \\
 when \quad b \leq r \leq c   \nonumber
\quad .
\end{eqnarray}

\begin{figure}
\includegraphics[width=12cm,angle=0]{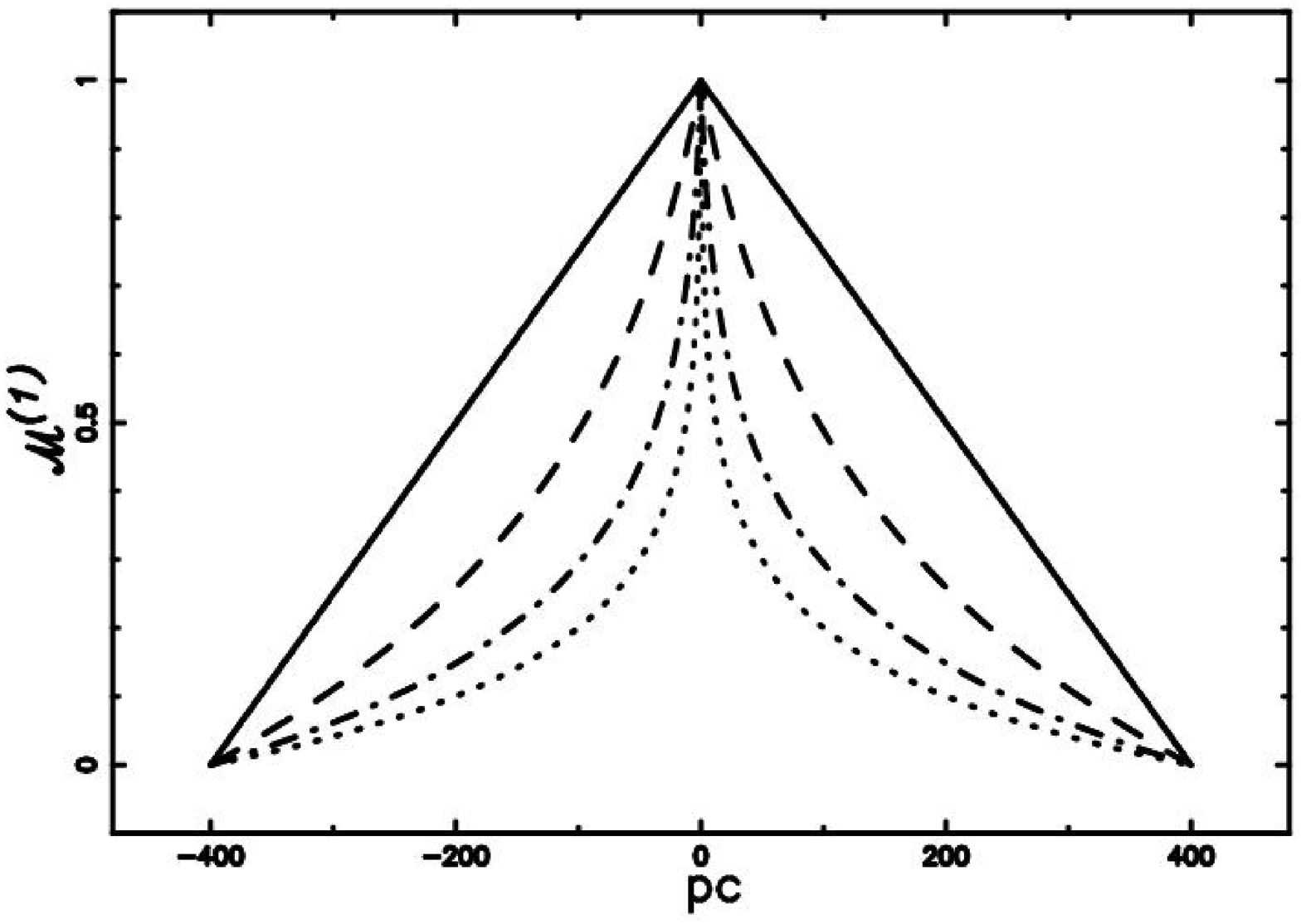}
\caption
{
Values of ${\mathcal M}^{(1)}$ or concentration  computed with
equations~(\ref{cvarab})    and (\ref{cvarbc})
when a =-400~pc, b=0~pc and c=400~pc :
$D(c) /D(b)= 1.0001 $  ( full line ),
$D(c) /D(b)= 10   $    ( dashed    ),
$D(c) /D(b)= 100  $    ( dot-dash-dot-dash ),
$D(c) /D(b)= 1000 $    ( dotted    ).
\label{variable_1D}}
    \end{figure}

The solution is found to be
\begin{eqnarray}
\label {cvarab}
C(r) =
{\frac { \left( \ln  \left( -{\it a_D}\,a+1+{\it a_D}\,b \right) -\ln
 \left( -{\it a_D}\,r+1+{\it a_D}\,b \right)  \right) {\it C_m}}{\ln
 \left( -{\it a_D}\,a+1+{\it a_D}\,b \right) }} \\
when \quad a \leq r \leq b  \nonumber
\quad ,
\end{eqnarray}
and
\begin{eqnarray}
\label {cvarbc}
C(r)=
-{\frac {{\it C_m}\, \left( -\ln  \left( {\it a_D}\,c+1-{\it a_D}\,b
 \right) +\ln  \left( {\it a_D}\,r+1-{\it a_D}\,b \right)  \right) }{
\ln  \left( {\it a_D}\,c+1-{\it a_D}\,b \right) }} \\
 when \quad b \leq r \leq c   \nonumber
\quad .
\end{eqnarray}
The new 1D theoretical solution
when  the  diffusion coefficient is variable is 
reported in Figure~\ref{variable_1D}.

Confront our Figure~\ref{variable_1D} with Figure~8 
in \cite{Schlickeiser2005} , where the density of primary 
CR is a function of the distance from the Galactic plane 
and different physical parameters.

\subsection{Fourier Series }

\begin{figure}
\vspace {100pt}
{\caption[]
{
Profile of ${\mathcal M}^{(3)}$ versus the number of sites as given by  
Monte Carlo simulation with regular lengths step 
 (filled circles ) 
 +  solution  of the mathematical diffusion 
(full line ) 
 + Fourier expansion 
(dot-dash-dot-dash line  )
. 
The parameters are 
NDIM=101 ,  L=25,
NTRIALS=10 000 ,
$m$, $l$ , $j$ ranging from $1$ to $21$.
}
\includegraphics[width=12cm,angle=0]{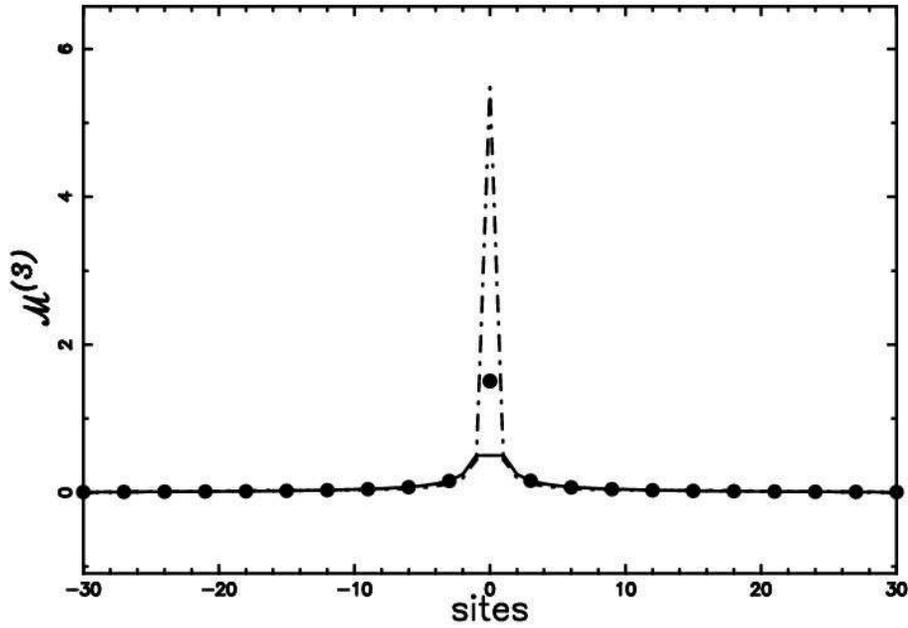} 
\label{iperbole_3d}}
    \end{figure}

In the case of restricted random walk the most straightforward way 
to compute the mean number of visits is 
via the solution of the Fokker-Planck equation,
which  in the case of simple, symmetric random walk 
equation reduces to the diffusion equation~\cite{Feller_2006}. 
Given the dimension  d, the random walk takes place on a finite domain  
$D=[-L/2,L/2]^d \subset R^d$, and  the boundaries are supposed  
 to be absorbing.
In this case the solutions of the diffusion equation 
can be obtained 
with the usual
methods of eigenfunction expansion \cite{Feller_2006,Morse1953}  and 
they are  cosine  series 
whose coefficients decay exponentially, see  equation~(25)
in \cite{ferraro2004}.
The  solutions particularized as a cuts  along one axis 
are  
\begin{equation}
{\mathcal M}^{(1)}(x) = \frac{2}{L}\sum_{m=0}^{\infty}\kappa_{m}
\cos\left[\frac{(2m+1)\pi x}{L}\right],
\label{eq:1d}
\end{equation}
\begin{equation}
{\mathcal M}^{(2)}(x,0)=\left(\frac{2}{L}\right)^2
\sum_{m,l=0}^{\infty}\kappa_{ml}
\cos\left[\frac{(2m+1)\pi x}{L}\right]
\label{eq:2d}
\quad ,
\end{equation}

\begin{equation}
{\mathcal M}^{(3)}(x,0,0)=\left(\frac{2}{L}\right)^3\sum_{m,l,j=0}^{\infty}\kappa_{mlj}
\cos\left[\frac{(2m+1)\pi x}{L}\right],
\label{eq:3d}
\end{equation} 

with  

\begin{equation}
a=\frac{L}{2}
\nonumber 
\quad ,
\end{equation}

\begin{equation}
\kappa_m=\frac{\left(2L^2 \right)}{(2m+1)^2\pi^2}
\nonumber 
\quad ,
\end{equation}

\begin{equation}
\kappa_{ml}=\frac{\left(2L^2 \right)}{\left 
((m+1)^2+(l+1)^2 \right)\pi^2}
\nonumber 
\quad ,
\end{equation}

\begin{equation}
\kappa_{mlj}=\frac{\left(2L^2 \right)}{\left 
((m+1)^2+(l+1)^2+(j+1)^2 \right)\pi^2}
\nonumber 
\quad .
\end{equation}
These three new solutions are the Fourier counterpart
 of the solutions of the mathematical
diffusion 
(\ref{cbc_1d}), (\ref{cbc_2d}) and  (\ref{cbc})  respectively.
The  comparison between solution in 3D  with methods of eigenfunction expansion,
mathematical  diffusion and Monte Carlo simulation   is 
carried out in Figure~\ref{iperbole_3d}.

\subsection{Levy random walks in 1D}
\label{levysec}
\begin{figure}
\includegraphics[width=12cm,angle=0]{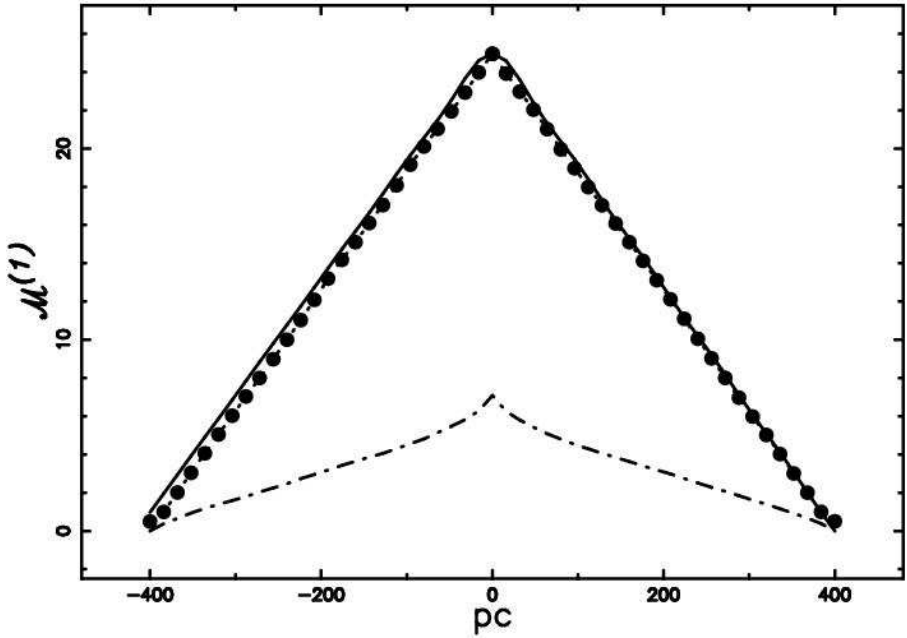}
\caption
{
Profile of ${\mathcal M}^{(1)}$  versus the distance from
the Galactic plane  as given by  
Monte Carlo simulation with regular lengths step 
  + theoretical results given by equation~(\ref{eq:1d})
 + 
Monte Carlo simulation with Levy random walk
(filled circles , dotted line  and  
dot-dash-dot-dash line respectively ). 
The parameters are 
NDIM=51 , L=25, 
$m$ ranging from $1$ to $20$, side =800~pc ,
NTRIALS=10000  and $\alpha$=1.8~. 
}
\label{cut1d}
\end{figure}

\begin{figure}
\includegraphics[width=12cm,angle=0]{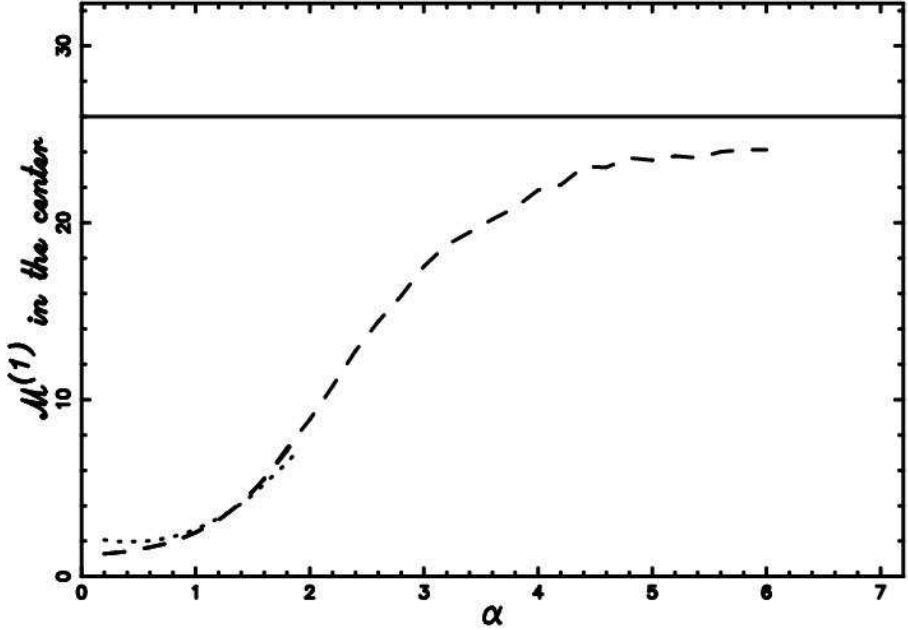}
\caption
{
Behaviour of ${\mathcal M}^{(1)}$   at the center of the lattice
as a function of  $\alpha$ (dashed--line) + 
theoretical solution (dotted--line) + asymptotic
behaviour at  (NDIM+1)/2  (full line).
Parameters as in Figure~\ref{cut1d}~.
}
\label{max}
    \end{figure}

In the case of the Levy~\cite{Levy1937} random walk in 1D
it should noted that 
$\delta x$, i.e. the single step, can be written as 
\begin{equation}
\delta x=\xi l,
\label{eq:deltax}
\end{equation}

where $l>0$ is the length of the  flight and $\xi$ is 
a stochastic variable taking values  $1$ and $-1$ with equal probability.
For the distribution of $l$ chose 
\begin{equation}
p(l) =C l^{-(\alpha+1)}
\quad ,
\label{pl}
\end{equation}
where  $0 < \alpha < \infty$, and then 
\begin{equation}
p(\delta x)=1/2p(l)
\label{eq:ptot}
\quad .
\end{equation}
Let  ${\mathcal S}^\alpha$  denote the average number of visits to 
a site obtained via a Monte Carlo simulation ; 
when  ${\mathcal M}^\alpha$ denotes  the corresponding values 
given by the equation~\cite{Gitterman2000}, 
\begin{equation}
{\mathcal M}^{\alpha}(x)=
\frac{2}{L}\sum_{n=1}^{\infty}\frac{L^\alpha}{(n \pi)^\alpha D_\alpha}
  \sin\left (\frac{m\pi(x+a)}{L} \right ) \sin\left 
(\frac{m\pi a}{L} \right )
\quad ,
\label{eq:final}
\end{equation}
where 
\begin{equation}
\label{eq:coeff}
D_{\alpha}=\frac{4}{\pi T} \left (\frac{L}{\pi} \right)^{\alpha}
\sum_{m=1}^{\infty}\frac{(-1)^m}{(2m+1)^{1+\alpha}}
\quad ,
\end{equation}
being   
\begin{equation}
\label{eq:tmed}
T=\sum_{x=0}^L {\mathcal S}^{\alpha}(x)
\quad  .
\end{equation}
Once we calculate the constant in pdf~(\ref{pl})  (probability density function)
 the  Pareto distribution, P, is obtained \cite{evans}:
\begin {equation}
P(l;a,c) = \frac {c a^c}{l^{c+1}} \quad ,
\label{pareto}
\end {equation}
where $  a \leq l < \infty $ , $ a~>0$ , $ c~>0$.
The average value is    
\begin{equation}
\overline {l}=
\frac {ca} {c-1} 
\quad ,
\end{equation}
which is defined for $c >1$,
and the variance  is 
\begin{equation}
\sigma^2 =
{\frac {{a}^{2}c}{ \left( -2+c \right)  \left( -1+c \right) ^{2}}}
\quad ,
\end{equation}  
which is defined for $c >2$. 
In our case the steps  are comprised between  
1 and  $l_{max}$ and the following pdf  , named Pareto truncated $P_T$ ,
has been used
\begin {equation}
P_T(l;l_{max},\alpha ) = \Bigr (
                    \frac{\alpha} {1 - \frac{ 1}{l_{max}}} 
                         \Bigl )     
\frac {1}{l^{\alpha +1}} \quad .
\label{eq:pdf}
\end {equation}
The  average value of $P_T$  is  
\begin{equation}
\overline {l} = 
-{\frac {\alpha\, \left( -{{\it l_{max}}}^{\alpha}+{\it l_{max}} \right) }{
 \left( {{\it l_{max}}}^{\alpha}-1 \right)  \left( -1+\alpha \right) }}
\quad ,
\end {equation} 
and the variance of $P_T$ is 
\begin {eqnarray}
 \sigma^2 =
{\frac {\alpha\, \left( -{{\it l_{max}}}^{2\,\alpha}+{{\it l_{max}}}^{\alpha
}-2\,{{\it l_{max}}}^{\alpha+2}\alpha-2\,{{\it l_{max}}}^{\alpha+1}{\alpha}^
{2}+{\alpha}^{2}{{\it l_{max}}}^{\alpha+2} \right) }{ \left( -2+\alpha
 \right)  \left( -1+\alpha \right) ^{2} \left( -{{\it l_{max}}}^{2\,
\alpha}+2\,{{\it l_{max}}}^{\alpha}-1 \right) }}
+            \nonumber  \\  
+{\frac {\alpha\, \left( {{\it l_{max}}}^{\alpha}{\alpha}^{2}-2\,\alpha\,{
{\it l_{max}}}^{\alpha}+4\,{{\it l_{max}}}^{\alpha+1}\alpha-{{\it l_{max}}}^{2}
+{{\it l_{max}}}^{\alpha+2} \right) }{ \left( -2+\alpha \right)  \left( -
1+\alpha \right) ^{2} \left( -{{\it l_{max}}}^{2\,\alpha}+2\,{{\it l_{max}}}
^{\alpha}-1 \right) }}
\quad . 
\end {eqnarray}
This variance  is new formula and is always defined for every value of $\alpha >$ 0;
conversely the variance of the 
Pareto distribution  can be defined  only when  $\alpha >$ 2.
From a numerical point of view a  discrete distribution of  
step's length 
is given by the following  pdf  
\begin{equation} 
p(l) dl   \propto  l^{-(\alpha+1)}~dl 
\quad ,
\label{distribution}
\end {equation}
and the set of lengths 
could be  obtained through a numerical computation
of the inverse function~\cite{Brandt1998}.
The interval of existence of the length-steps 
is  comprised between 1  (the minimum length) 
and $l_{max}$=NDIM    that is  number 
that characterises the lattice~.
We can  now examine how the  impact  of the Levy random walk
influences the results of the  simulations in 1D;
in the following NTRIALS will represent the number of different
trajectories implemented  in the Monte Carlo simulations.
The memory grid ${\mathcal M}^{(1)}$ as a function of the distance 
from the center takes the curious triangular shape 
visible in Figure~\ref{cut1d} ;
a theoretical explanation  resides is the
random walk  in bounded 
domains \cite{Gitterman2000,Ferraro_2006}.
In that figure the theoretical 
solution ( formula~(25)
in~\cite{ferraro2004} when dimension d=1)
is also reported
and  the  fit is  satisfactory.
In Figure~\ref{cut1d}  we  report the visitation grid 
as given by the presence of the Levy random walk 
once a certain value of $\alpha$ is chosen; is evident the effect
of lowering the value of the visitation grid 
respect to the regular  random walk.
The effect of increasing  the exponent $\alpha$ 
that  regulates the lengths of the steps could be summarised 
in Figure~\ref{max} in which we report 
the value of the memory grid  ${\mathcal M}^{(1)}$ 
at the center of the grid. From the previous figure 
is also  clear the transition from Brownian motion 
($\alpha \approx  2$ ) 
 to regular    motion ($\alpha \approx  10$ ). 
\subsection{Levy diffusion coefficient in 1D}

\label{diff_coef}
\begin{figure}
\includegraphics[width=12cm,angle=0]{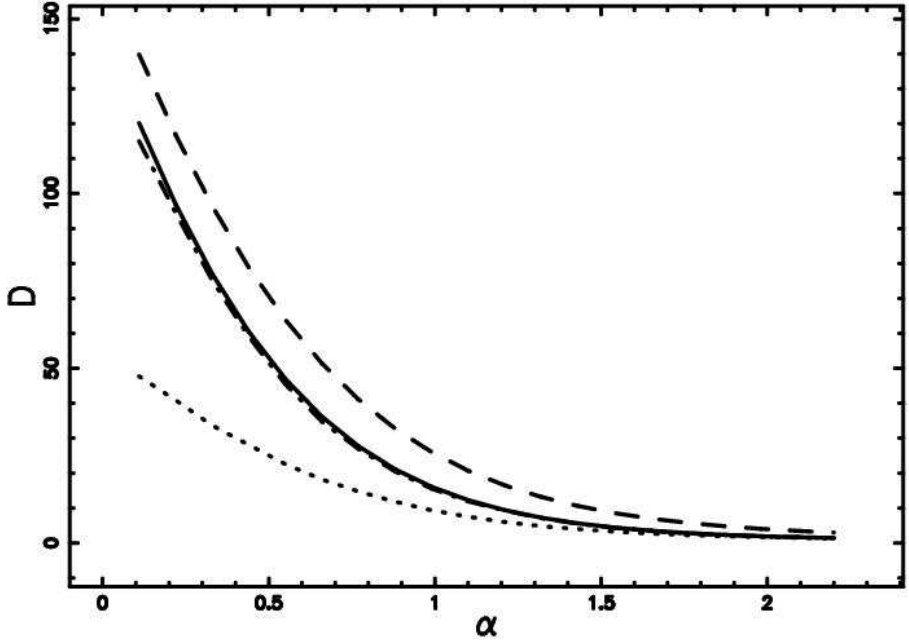} 
\caption
{
Behaviour of the diffusion coefficient 
as function  $\alpha$:
$D_{theo}^I   $  ( full line ), 
$D_{theo}^{II}$  ( dashed ), 
$D_{num }^{I}$ ( dot-dash-dot-dash ),
$D_{num }^{II}$( dotted ).
}
\label{coeff_1d}
    \end{figure}

The 1D diffusion coefficient in presence of 
Levy flights can be derived from analytical or
numerical arguments that originate  four different definitions
summarised in Figure~\ref{coeff_1d}.
Let the single displacement,$\delta x$, be defined by
\begin{equation}
\delta x=\xi l,
\label{eq:deltax2}
\end{equation}
where  $l>0$  is the length of the flight and $\xi$ is
a discrete random variable characterised by the following
probabilities p(1)=+1 and p(2) =-1.
The probability $p(\delta x)$ of a single displacement
 $\delta x$
is
\begin{equation}
p(\delta x)=\frac{1}{2}p(l) \quad .
 \label{eq:probdx}
\end{equation}
This is the product of the probability $p(\xi)=1/2$ to go at
right or at left by  the probability
 $p(l)$ to make a step of length  $l$.
From  $\langle\delta x\rangle=0$ and from the independence of the
 little steps follows
\begin{equation}
\langle x(n)\rangle=n\langle\delta x\rangle=0 \quad  ,
\end{equation}
where  $x(n)$  is the walker's position at the time  $n$ ( after
$n$ steps). From  $\langle\delta x\rangle=0$ follows
\begin{equation}
 \langle\delta^2 x\rangle=\sigma^2(\delta x), \quad \langle x^2(n)\rangle=\sigma^2(x(n)),
\label{eq:primvar}
\end{equation}
and due to the independence of the small steps ,
\begin{equation}
\sigma^2(x(n))=n \sigma^2(\delta x)
\label{eq:secvar}
\quad .
\end{equation}
The connection with the diffusion coefficient is through the
theoretical relationship
\begin{equation}
\label{eq:diff}
D=\frac{1}{2} \sigma^2(\delta x),
\end{equation}
thus
\begin{equation}
\label{eq:scdf}
D=\frac{1}{2n} \sigma^2( x)
\quad  .
\end{equation}
This formula can be compared with the well known formula that can
be found on the textbooks , see for example formula~(12.5)
in~\cite{gould}. We should now compute $\sigma^2(\delta x)$. Since
$\langle\delta x\rangle=0$ and $\xi$, $l$ are 
random independent variables we have
\begin{equation}
\sigma^2(\delta x)=\langle\delta^2 x\rangle=\langle(\xi
l)^2\rangle=\langle\xi^2\rangle\langle l^2\rangle
 \label{eq:comp}
\quad .
\end{equation}
Since  $\langle\xi^2\rangle=1$
\begin{equation}
\sigma^2(\delta x)=\langle l^2\rangle \label{eq:finp}
 \quad .
\end{equation}
We can continue in two ways .
A  first one consists into  evaluate $l^2$ according to the 
discrete distribution represented by 
equation~(\ref{distribution}).
This originates the first theoretical diffusion coefficient ,
in the following $D_{theo}^I$.
A second way consists into evaluate 
 $\langle l^2\rangle$ in view of 
pdf~(\ref{eq:pdf})  
\begin{equation}
\langle l^2\rangle=
-{\frac {\alpha\, \left( -{{\it l_{max}}}^{\alpha}+{{\it l_{max}}}^{2}
 \right) }{ \left( {{\it l_{max}}}^{\alpha}-1 \right)  \left( -2+\alpha
 \right) }}
\label{eq:lquadro}
 \quad ,
\end{equation}
and consequently ,
\begin{equation}
D_{theo}^{II}=
1/2\,{\frac {\alpha\, \left( {{\it l_{max}}}^{\alpha}-{{\it l_{max}}}^{2}
 \right) }{ \left( {{\it l_{max}}}^{\alpha}-1 \right)  \left( -2+\alpha
 \right) }}
\quad ,
\label{d_theo}
\end{equation}
where the new $D_{theo}^{II}$ is the second theoretical diffusion coefficient.

The  numerical techniques  allow to implement two different definitions
of the diffusion coefficient.
The first one considers an infinite lattice and a variable number 
of Monte Carlo steps N (30,31...35) each one  connected with an index j.
The diffusion coefficient in then computed as :
\begin{equation}
D_{num}^{I} = \frac { \sum _{i=1} ^{6} { \frac {R(i)^2}{2N(i)} }}{6}
\quad .
\end{equation} 
The second  evaluation  implements 
a typical Monte Carlo run on a lattice , see for example the data
in Figure~\ref{cut1d}. 
Now $R^2$ is fixed and takes the value $\frac{NDIM+1}{2}$  
in contrast with N , the number of iterations before to reach the
two boundaries, that is different in each run.
We therefore have
\begin{equation}
D_{num}^{II} = \frac { R^2}{2\overline{N}} 
\quad ,
\end{equation} 
where $\overline{N}$ is the average number of iterations 
necessary to reach the external boundary in each trial.
Up to now in order to test the value  of the  diffusion coefficient
we have fixed in one the minimum value that the step can take.
We now explore the case in which the minimum 
length in  pdf~(\ref{eq:pdf})  is $l_{min}$ .
On adopting  the same technique that 
has lead to deduce $D_{theo}^{II}$   we obtain
a new  more general expression , $D_{theo}^{II,G}$ ,
for the 1D   diffusion coefficient  
\begin{equation}
D_{theo}^{II,G} =
-1/2\,{\frac {\alpha\, \left( {l_{{\min}}}^{2}{l_{{\max}}}^{\alpha}-{l
_{{\max}}}^{2}{l_{{\min}}}^{\alpha} \right) }{ \left( \alpha-2
 \right)  \left( -{l_{{\max}}}^{\alpha}+{l_{{\min}}}^{\alpha} \right) 
}}
\label{DLevy}
\quad ,
\end{equation}
or  introducing  the ratio $r_l$ = $\frac{l_{min}}{l_{max}}$ 
\begin{equation}
D_{theo}^{II,G} =
1/2\,{\frac {\alpha\,{l_{{\max}}}^{2} \left( -{r_{{l}}}^{2}+{r_{{l}}}^
{\alpha} \right) }{ \left( \alpha-2 \right)  \left( -1+{r_{{l}}}^{
\alpha} \right) }}
\label{DLevy2}
\quad .
\end{equation}

\section{Energy evaluations}
\label{physics}
In order to introduce  the  physical 
mean free path  we should remember that an important
reference quantity is the
 relativistic ions  gyro-radius, $\rho_Z$.
Once the energy  is expressed in $10^{15}$eV
units ( $E_{15}$) , and the magnetic field in $10^{-6}$ Gauss ( $H_{-6}$)
we have:
\begin {equation}
\rho_Z = 1.08   \frac {E_{15}}  {H_{-6} Z  } pc
\label{rho}
\quad ,
\end   {equation}
where Z is  the atomic number.
The solutions of the mathematical diffusion 
in presence of the steady state do not require 
the concept of mean free path.
The solutions of the diffusion equation trough a Fourier expansion,
Levy flights and Monte Carlo methods conversely  require that 
the mean free path  should be specified like a fraction 
of the size of the box in which the phenomena is analysed~.
The presence of irregularities allows   the diffusion  
of the charged particles. 
The mean free path $\lambda_{sc}$ ,
see for example ~\cite{longair} , 
is  
\begin{equation}
\lambda_{sc} \approx  \rho_Z \Phi^{-2}
\quad ,
\end{equation}
with  $\Phi=\frac {B_1}{B_0}$  where  ${B_1}$ are ${B_0}$
are the strength of the random and mean magnetic field respectively.
The type of adopted turbulence, weak or strong, the spectral
index of the turbulence determine the exact correspondence 
between energy and mean free path.
Some evaluations on the relativistic  mean free path of cosmic rays 
and Larmor radius can be found  in  
\cite{longair,Jokipii_1973,Wentzel_1974,Sigl_2000}.
The transport of  cosmic rays 
with the length of the step
equal to the relativistic ion gyro-radius is  
called Bohm diffusion \cite{Bohm}
and   the diffusion coefficient will be energy dependent.
The assumption of the Bohm diffusion allows to fix a one-to-one 
correspondence between energy and length of the step
of the random walk. 
The presence of a short wavelength drift instability
in the case of quasi-perpendicular shocks  
drives the  diffusion coefficient to the 
"Bohm" limit \cite{Zank_1990,Jones_1991}.
The maximum energies that can be extracted from SNR and
Super-bubbles are now introduced.
The transit times in the halo of our galaxy are computed in the
case of a fixed  and variable magnetic field.
The modification  of the  canonical
energy probability density function  ,
p(E)~$\propto~E^{-2}$,
due to an  energy dependent diffusion coefficient is outlined.

\subsection{Maximum available energies}

The maximum energy available in the accelerating regions is connected
with their  size,
in the following $\Delta~R$  and with the Bohm diffusion 
\begin{equation}
\Delta~R = \rho_Z 
\quad .
\label{maximum_energy}
\end{equation}

The most promising sites are  the  SNR as well as the super-shell.
In the case of SNR, the analytical solution
for the radius can be found in~\cite{Dalgarno1987}~(equation~10.27) :
\begin{equation}
R(t)= \left ( \frac {25 E^{expl} t_{SNR}^2 }
              {4 \pi \rho_0 } \right  )^{1/5}
\label{eq:radiussnr}
\quad ,
\end{equation}
where $\rho_0$  is the density of the surrounding  medium
supposed to be constant , $E^{expl}$ is the energy of
the explosion,
and  $t_{SNR}$ is   the age  of the SNR.
Equation~(\ref{eq:radiussnr})  can be expressed by  adopting
the astrophysical units
\begin {equation}
R =  (\frac {E_{51}^{expl} t_4^2  }  {n_0})^{1/5} 12.47 \mathrm{pc}
\quad,
\end  {equation}
here
$t_4$  is   $t_{\mathrm{SNR}}$ expressed   in units of $10^4$ \mbox{years},
$E_{51}^{expl}$ is the energy expressed  in units of $10^{51}$ \mbox {ergs} and
$n_0$ is the density  expressed in $\mathrm{particles~}$ ${\mathrm{cm}^{-3}}$~.
In agreement with~\cite{Dalgarno1987}
$\rho_0$ = $n_0$ m  was taken,  where m = 1.4 $m_H$~:
this energy conserving phase ends at $t_4 \approx 1.4 $.
The thickness of the advancing  layer  is  $\Delta~R$  $\approx$ $R/12$
according to~\cite{Dalgarno1987} and this
allows  us to deduce the maximum energy ,$E^{max}_{15}$,
that can be extracted from SNR for a relativistic ion , see 
formula~(\ref{maximum_energy}):
\begin {equation}
E^{max}_{15} =
0.961 \bigl  (\frac {E_{51}^{expl} t_4^2} {n_0} \bigr )^{1/5}  H_{-6} Z
\quad .
\end{equation}
This new formula is in agreement with  more detailed
evaluations~\cite{Lagage}.

The  super-shells   have been observed  as   expanding  shells,
or holes,  in the H\,I-column density   distribution  of  our galaxy
\cite{heiles}.
The dimensions of these objects  span from 100~\mbox{pc}
to 1700~\mbox{pc}
and often present  elliptical shapes or elongated features,
that can be explained  as an  expansion in  a
non-homogeneous  medium~\cite{Zaninetti2004}.
An explanation of the  super-shell
is obtained from the theory of the Superbubbles , in the following SB,
whose  radius is ~\cite{Dalgarno1987}
\begin {equation}
R=\left[\frac{25}{14\pi}\right]^{1/5}
  \left(\frac{E_0R_{\mathrm{SN}}}{{\rho}}\right)^{1/5} t^{3/5}
 ,
\label{eq:firstradius}
\end {equation}
where  t is the considered time,
$R_{\mathrm{SN}}$ is the rate of supernova explosions, and 
$E_0$ the energy of each supernova;
this equation should be considered valid only
if  the altitude of
the OB associations from the Galactic ,$z_{OB}$=0, is zero,
and  only  for propagation  along the Galactic plane.

When the astrophysical units are adopted
the following is found
\begin{equation}
R =111.56\;  \mathrm{pc}(\frac{E_{51}^{expl} t_7^3 N^*}{n_0})^{\frac {1} {5}}
,
\label{eqn:raburst}
\end{equation}
where
$t_7$             is the time  expressed  in $10^7$ \mbox{yr} units,
$E_{51}^{expl}$   is the  energy in  $10^{51}$ \mbox{erg},
$n_0$      is the density expressed  in particles~$\mathrm{cm}^{-3}$,
and $N^*$  is the number of SN explosions
in  $5.0 \cdot 10^7$ \mbox{yr}.
As  an example,  on inserting in
equation~(\ref{eqn:raburst})
the typical parameters of
GSH~238~\cite{heiles} ,
$E_{51}^{expl}$=1       ,
$t_7$=2.1 , $n_0$=1 and  $N^*$=80.9~, 
the radius turns out to be 419~pc in rough agreement
with the average radius of GSH~238 ,  $\approx$ 385~pc.

The maximum energy ,$E^{max,sb}_{15}$, 
that can be extracted from a SB  ( from formula~(\ref{maximum_energy}))
in the Bohm diffusion is therefore:
\begin {equation}
E^{max,sb}_{15} =
8.61 \bigl  (\frac {E_{51}^{expl} N^*} {n_0} \bigr )^{1/5} t_7^{3/5}  H_{-6} Z
\quad.
\label{emaxsb}
\end{equation}
Once the typical parameters of GSH~238 are inserted
in equation~(\ref{emaxsb}) we obtain
$E^{max,sb}_{15}=323.6~H_{-6}~Z$. 
The exact value of the magnetic
field  in the expanding layer of the SB is still subject 
of  research , the interval   $H_{-6}~\in \{1,...200\}$
being generally accepted~\cite{Parizot}.
The interested reader can compare our equation~(\ref{emaxsb})
with equation~(15) in~\cite{Parizot}.

\subsection{Transit times}

The absence of the isotope  $~^{10}Be$,
which has a characteristic  lifetime
$\tau_r=  0.39 \times 10^{7}~yr$ ,
implies   the following inequality on $\tau$,
the mean cosmic ray residence time ,
\begin{equation}
\tau >  \tau_r
\quad .
\label{berillium}
\end{equation}

Given a box of side L with   a proton inserted
at the center,  the classical time of crossing
the box  $t_c$  is :
\begin{equation}
t_c  =\frac{L}{2c}
\quad  .
\label {tc}
\end {equation}
Once  the  {\it pc} and {\it yr} units are chosen ,
$c=0.306  \frac {pc}{yr}$  and
\begin{equation}
t_c  =1.63  L_{1}~yr
\quad  ,
\end{equation}
where $L_1$ is the side of the box expressed in 1pc units.

When the side of the box is $L_1$=800, 
$t_c  \approx 1304  ~yr$,
and   $t_c \ll \tau_r$, we should discard  the diffusion of the cosmic rays
through  the ballistic model.

We now  analyse firstly the case in which the value of the diffusion
coefficient is constant and secondly the case in which the diffusion
coefficient is variable.

\subsubsection{Transit times when D is fixed}

\begin{figure}
\includegraphics[width=12cm,angle=0]{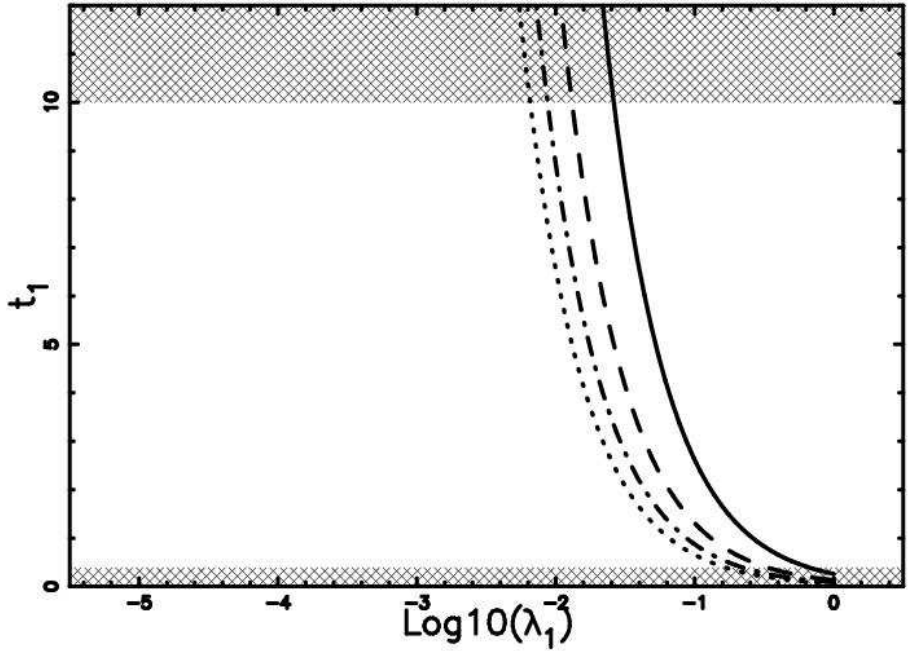}
\caption
{
Transit times expressed in $10^7$~yr units
as a function  $\lambda_1$ when
$L_1$=800  :
$c_{tr}=0.2 $  ( full line),
$c_{tr}=0.4 $  ( dashed),
$c_{tr}=0.6 $  ( dot-dash-dot-dash),
$c_{tr}=0.8  $  ( dotted).
The forbidden  region is represented
through  cross-hatched lines.
}
\label{times}
    \end{figure}

From  equations~(\ref {r2_continuum}) and
(\ref{coefficient})  we easily  find
the time , t, necessary  to travel the distance
R
\begin{equation}
t=\frac {R^2}  {  c_{tr}  c  \lambda}
\quad ,
\end{equation}
where  $v_{tr}=c_{tr}~c$  .
We continue inserting $R=\frac{L}{2}$ , $L=L_1~pc$
and  $t=t_1~10^7~yr$  ,
\begin{equation}
t_1=0.815 \times 10^{-7} \frac {L_1^2}  { c_{tr}    \lambda_1}
\quad .
\label{t1_senza}
\end{equation}
Figure~\ref{times}  reports a plot of the transit
times versus the two main parameters $\lambda_1$ and $c_{tr}$
together with the constrains represented by  the
lifetime  of the $~^{10}Be$. In this case the lifetime
of the phenomena ( cross-hatched region ) is 
tentatively fixed in $t_1=10$.

Conversely in presence of Levy flights  
(diffusion coefficient as  given by formula~\ref{DLevy2}) 
the transit time  $t_1^L$   
takes the form  
\begin{equation}
t_1^L=-0.815 \times 10^{-7} \frac {L_1^2}  { c_{tr}    \lambda_1}
{\frac { \left( \alpha-2 \right)  \left( -1+{r_{{l}}}^{\alpha}
 \right) }{\alpha\, \left( {r_{{l}}}^{2}-{r_{{l}}}^{\alpha} \right) }}
\quad .
\label{t1_senzal}
\end{equation}
The effect of the Levy flights is to increase  the value of 
the transit times.
Is interesting to realize that, in  this astrophysical 
form of the Levy flights,  the steps  with  transport velocity
greater than the light velocity are avoided.

\subsubsection{Transit times when D is variable}

\label{dvariable}
The  value of the  pressure in  the Solar surroundings
takes the value , $ p_{12} 10^{-12} {\mathrm {dyne~cm}}^{-2}$
\cite{boulares}
when  the pressure   is expressed in $10^{-12}{\mathrm {dyne~cm}}^{-2}$
units , $p_{12}$.

\begin{figure}
\includegraphics[width=12cm,angle=0]{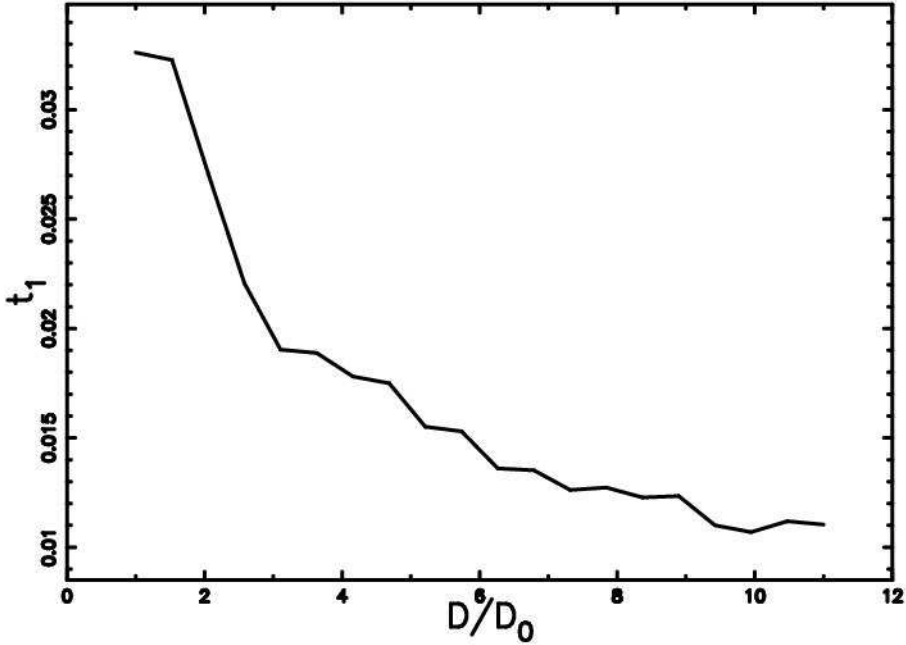}
\caption
{
Transit times  expressed in $10^7$~yr units
as a function of  $D(400pc) /D(0)= 10$ 
 when
$L_1$=800,
$\lambda_1$=8 and
$c_{tr}=0.2 $.
Assuming Z=1 ,$H_{-6}$=1 we have ( Bohm diffusion) $E_{15}$=7.38
}
\label{times_v}
    \end{figure}

Assuming equipartition between thermal pressure and magnetic pressure,
the magnetic field $B_0$ at the Galactic plane turns out
to be
\begin{equation}
B_0 = \sqrt  5 10^{-6}{p_{12}} ~Gauss
\quad  .
\end{equation}
The behaviour of the magnetic field with the Galactic height
z in pc is  assumed  to vary  as
\begin{equation}
B = \frac{B_0} {1+a_D z}
\quad ,
\end{equation}
with  $D_0$   defined  in Section~\ref{sec_diffvar} through 
the Bohm diffusion ,
\begin{equation}
D_0 = \frac{1}{d} c_{tr} \frac {E_{15}} {H_{-6} Z} 0.166~ 10^7~~
\frac{pc^2}{(10^7 yr)}
\quad .
\end{equation}
In order to see how the presence of a variable diffusion
coefficient modifies the transit times given by
equation~(\ref{t1_senza}) we performed the following Monte Carlo
simulation.
The 1D random walk    has  now the length of the step function of the
distance from the origin.
In order to mimic equation~(\ref{D_variable}) the step has
length
\begin{equation}
\lambda =\lambda1 (1+ a_D z )
\quad.
\label{l_variable}
\end{equation}
Typical behaviour of the influence of the variable D
on  transit times has been reported in
Figure~\ref{times_v}.
It is clear that an increasing D ( a decreasing B) lowers
the transit times; a rough evaluation predicts
that  transit times decrease $\propto~D(400pc) /D(0)$.

\subsection{The spectral index}

At present  the observed differential spectrum of cosmic rays
scale as $E^{-2.75}$ in the interval $10^{10}~eV-5.0~10^{15}~eV$
and after $5~10^{15}~eV$ as  $E^{-3.07}$.

Up to now the more interesting result on the prediction
of the spectral index resulting from particle acceleration
in shocks , i.e. SNR or SB , is due to \cite{Bell_II,Bell_I,longair}.

The predicted differential energy spectrum of the high energy
protons is
\begin{equation}
N(E)dE \propto E^{-2} dE
\quad .
\end {equation}

This is the spectral index
where the protons are accelerated ; i.e. the region
that has size  $\approx~\frac {R}{12}$ , where R is the radius
of the SNR or SB.  Due to the spatial diffusion,
the spectral index changes.  The first assumption is that
the transport occurs at lengths of the order of magnitude
of the gyro-radius, the previously introduced 
Bohm diffusion. 
Two methods are now suggested 
for the modification of the original index 
as due to the diffusion :
one  deals
with the concept of advancing segment/circle  or sphere
and the second one is connected with the asymptotic behaviour
of the 3D concentration.
\subsubsection{Index from the diffusion coefficient}

On considering a  diffusion function 
of the dimensionality d  (see formula~(\ref{r2_continuum})) , 
the square of the distances in presence of Bohm diffusion will be
\begin{equation}
{\overline R^2}  =
\frac{ 0.33 10^7 c_{tr} E_{15} t_1 } {H_{-6} Z}~pc^2
\quad .
\end{equation}
Let us consider two energies $E_{15,1}$ and  $E_{15,2}$
with two corresponding numbers  of protons (Z=1) , $N_1$ and $N_2$.
We  have two expressions   for the two advancing spheres (d=3)
characterised by radius $R_1(E_{15,1})$   and $R_2(E_{15,2})$.

The ratio of the two densities
per unit volume , $N_{2,1}$, gives the differential spectrum
\begin{equation}
N_{2,1} = \frac {\rho_1 } {\rho_2 }=
\frac {N_2}{N_1} ( \frac {E_{15,1}}  {E_{15,2}})^{3/2}
\quad .
\end{equation}
But in the source
\begin{equation}
\frac {N_2}{N_1} =  ( \frac {E_{15,1}}  {E_{15,2}})^{2}
\quad ,
\end{equation}
and therefore
\begin{equation}
N_{2,1} \propto
( \frac {E_{15,1}}  {E_{15,2}})^{3.5}
\quad or
\propto E^{-3.5}
\quad ,
\end{equation}
when d=3  or 
\begin{equation}
N_{2,1} \propto  
 E^{-(2 +d/2)}
\quad ,
\end{equation}
when  the concentration is function of d.  
This new demonstration simply deals
with the concept of advancing segment/circle  or sphere.

\subsubsection{Index from the 3D concentration}

Starting from  the asymptotic behaviour  of the  3D concentration ,
formula~(\ref{approximate3D})  
the ratio of the two energy  concentrations is possible to derive
a new formula 
\begin{equation}
N_{2,1} \propto  
 E^{-(2 + 1)}
\quad .
\end{equation}

\section{CR Diffusion in the SB  environment}
\label{sb}
Once equipped to calculate the spatial diffusion and temporal evolution
of the CR  attention now moves to  the 3D evolution of CR once they have been
accelerated in the advancing layer of super-shells.

\subsection{SB}

The  super-shells   have been observed  as   expanding  shells,
or holes,  in the H\,I-column density   distribution  of  our galaxy
and
the dimensions of these objects  span from 100~\mbox{pc}
to 1700~\mbox{pc}~\cite{heiles1979}.
The  elongated shape of these
structures  is   explained through introducing
theoretical objects named  super-bubble (SB);
these  are created by  mechanical energy input from stars 
\cite{pikelner,weaver,Dalgarno1987}.

The evolution of the SB's in the ISM can be followed
once the following parameters are introduced
$t_\mathrm{\mathrm {age}}$ ,
the age of the SB in yr,
$\Delta t$,
the time step to be inserted in the differential equations,
$t^{\mathrm{burst}}$,
the time after which the bursting phenomena stops ,
$N^*$,  which  is the number of SN explosions
in  $5.0 \cdot 10^7$ \mbox{yr}.
An analytical solution of the advancing radius as
a function of time  can be found \cite{Dalgarno1987}
when the  ISM  has constant density.

The basic equation that governs the evolution
of the SB~\cite{Dalgarno1987}, \cite{mccrayapj87}
is  momentum conservation
applied to a pyramidal section,
characterised
by a solid angle, $\Delta \Omega_j$:
\begin {equation}
\frac{d}{dt}\left(\Delta M_j \dot{R}_j\right)=p R_j^2
\Delta \Omega_j
,
\label{eq_momentum}
\end {equation}
where   $\Delta \Omega_j$ is the solid angle along a given direction
and the mass
is confined into a thin shell with mass $\Delta M_j$.
The subscript $j$  indicates
that this is not a spherically symmetric system.

In our case  the density
is given  by the sum of three exponentials
\begin{equation}
n(z)  =
n_1 e^{- z^2 /{H_1}^2}+
n_2 e^{- z^2 /{H_2}^2}+
n_3 e^{-  | z |  /{H_3}}
\,  ,
\label{equation:ism}
\end{equation}
\label{Sec_ISM}
where   {\it z}  is the  distance  from  the Galactic plane in pc,
$n_1$=0.395 ${\mathrm{particles~}}{\mathrm{cm}^{-3}}$, $H_1$=127
\mbox{pc},
$n_2$=0.107 $\mathrm{particles~}{\mathrm{cm}^{-3}}$, $H_2$=318
\mbox{pc},
$n_3$=0.064 $\mathrm{particles~}{\mathrm{cm}^{-3}}$, and  $H_3$=403
\mbox{pc}
and this requires  that  the differential equations  should be
solved along
a given number of directions each denoted by  $j$ .
By varying the basic parameters,
which  are   the bursting time  and the
lifetime  of the source
characteristic structures   such as    hourglass-shapes,
vertical walls and  V-shapes,
can   be obtained.
The SB are then disposed on a spiral structure as given
by the percolation theory   and the  details can
be found in \cite{Zaninetti2004}.

The  influence of the Galactic rotation on the results 
can be be obtained by  introducing 
the  law  of the Galactic rotation 
as given by~\cite{Wouterloot_1990},
\begin{equation}
V_{\mathrm{R}} (R_0)  =220 ( \frac {R_0[\mathrm{pc}]} {8500})^{0.382} 
\mathrm {km~sec}^{-1}
,
\label {vrotation}
\end {equation} 
here  $R_0$ is the radial distance from the center of the Galaxy 
expressed in pc.
The original circular  shape of the superbubble  at z=0
transforms in an ellipse through  the 
 following 
transformation, $T_{\mathrm {r}}$,   
\begin{equation}
T_{\mathrm{r}} ~
 \left\{ 
  \begin {array}{l} 
  x\prime=x  +  0.264 y ~t\\\noalign{\medskip}
  y\prime=y\\\noalign{\medskip}
  z\prime=z,
  \end {array} 
  \right. 
\label{trotation}
\end{equation}
where {\it y} is expressed in pc and t in $10^7~{\mathrm{yr}}$ units
\cite{Zaninetti2004}.
In the same way the effect of the shear velocity 
as function of the distance  y  from the center of the expansion,
$V_{shift}(y)$  
can be easily obtained on Taylor expanding  equation~(\ref{vrotation})
\begin{equation}
V_{shift}(y) = 84.04  \frac {y}{R_0}
\, Km/s 
\quad .
\label{vshift}
\end{equation}
This is the amount of the shear velocity  function
of y but directed toward x   in a framework in which the
velocity  at the center of the expansion is zero and is new.

\subsection{Monte Carlo diffusion of CR from SB}

The diffusing algorithm  here  adopted is the 3D random walk
from many injection points ( in the following IP),
disposed on $N_{SB}$ super-bubbles.
In this case a time dependent situation is explored ;
this is  because the diffusion from the sources cannot be greater
than the lifetime of SB.
The rules are :
\begin{enumerate}
\item The first of $N_{SB}$ SB  is chosen.
\item The IP are generated  on the surface
      of each SB.
\item The 3D random walk starts from each IP with a fixed  length of a
      step. The number of visits is recorded on a 3D grid
      $\mathcal {M}$ .
\item After a given number of iteration , ITMAX , the process
      restarts from (iii) taking into account  another IP.
\item After the diffusion from the n SB  ,
      the process restarts  from (i) and the (n+1) SB
      is considered.
\end {enumerate}

\begin{figure}
\vspace {100pt}
\caption[]
{
Three dimensional surface representing the surface brightness, front view.
The ISO--density contours are fixed through   {\it coef=0.55}.
The number of super-bubbles $N_{SB}$ is 260  and the time evolution
is followed as in \cite{Zaninetti2004}~,
$\lambda_1$=172.94, IP=200 , ITMAX= 20000 ,
$t_D~=2.26~10^{7}$~yr    and
$c_{tr}=0.5 $.
Assuming Z=1 ,$H_{-6}$=1 we have ( Bohm diffusion) $E_{15}$=159.68~.
 }
\includegraphics[width=12cm,angle=0]{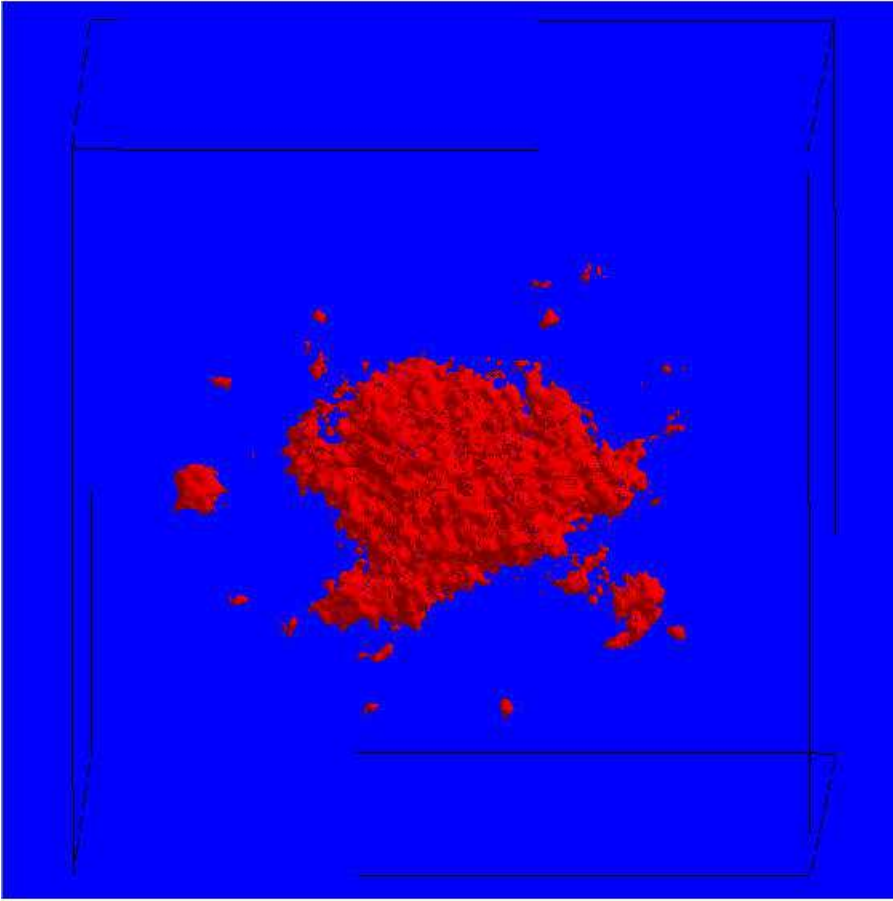}
\label{isosurf_plane}%
    \end{figure}

       The spatial displacement of the 3D grid
       ${\mathcal M}({i,j,k})$  can be visualised
        through the ISO-density
       contours. In order   to do so,
       the  maximum  value ${ {\mathcal M}(i,j,k)}_{max}$
       and the minimum  value ${ {\mathcal M}({i,j,k})}_{min}$~
       should be extracted
       from  the three-dimensional grid.
       A value  of the visitation or concentration grid can then be    fixed by
       using the following  equation:
\begin{equation}
{ {\mathcal M}({i,j,k})}_{chosen} =
{ {\mathcal M}({i,j,k})}_{min} +
{({{ {\mathcal M}({i,j,k})}_{max} -
{{\mathcal M}({i,j,k})}_{min}}) \times  {coef}}
\quad,
\end{equation}
where {\it coef} is a parameter comprised between
0 and 1.
This iso-surface rendering is reported in
Fig~\ref{isosurf_plane}~ ; the Euler
       angles characterising the point of view of the
       observer are also reported~.

\begin{figure}
\vspace {100pt}
\caption[]
{
The same as  Figure~\ref{isosurf_spiral}
but ITMAX= 600 ,
$t_D~=6.7~10^{5}$, and  {\it coef=0.40}.
}
\includegraphics[width=12cm,angle=0]{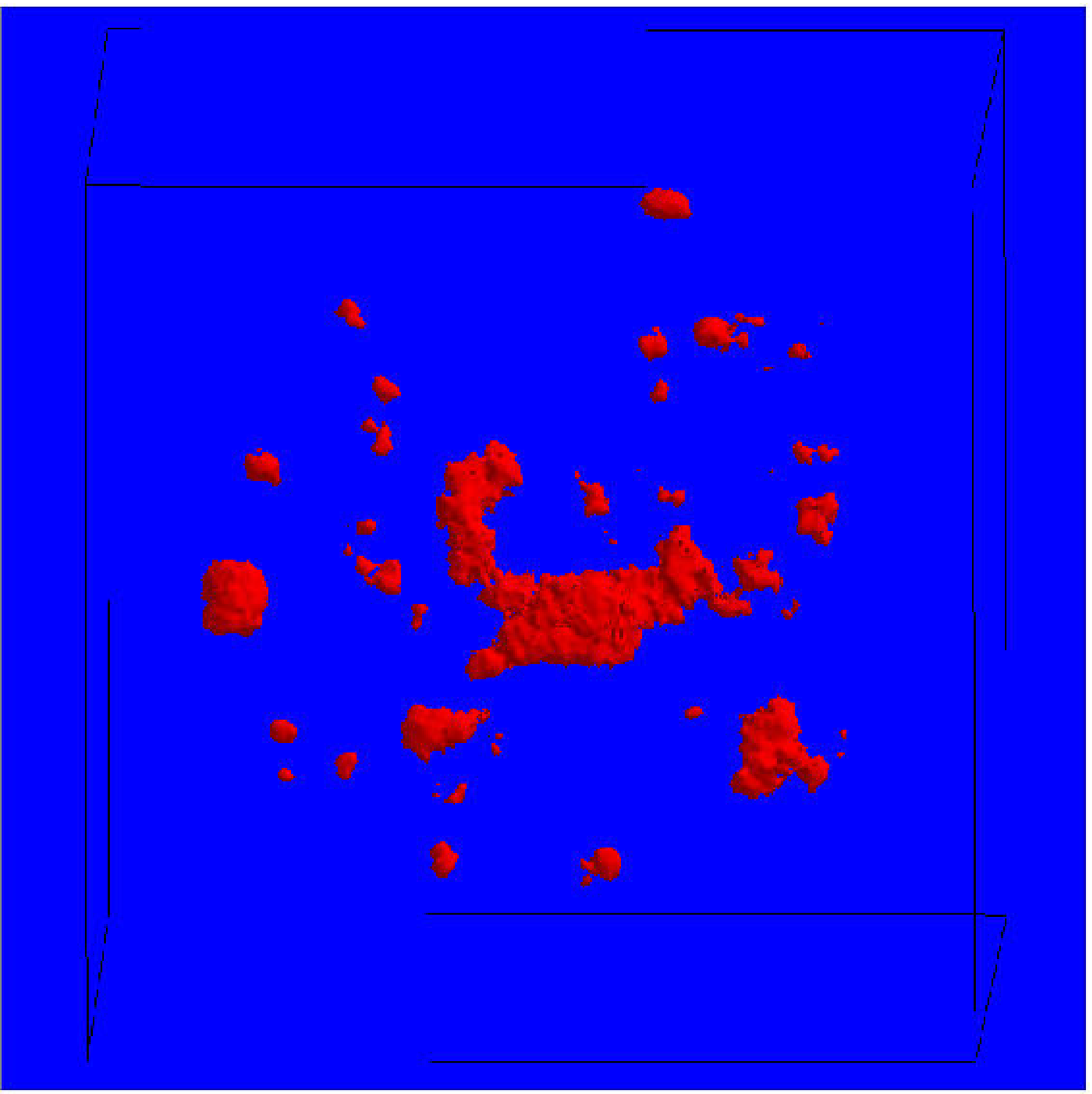}
\label{isosurf_spiral}%
    \end{figure}

From the previous figure it is possible to see  a progressive intersection
of the contributions from different super-bubbles:
once we decrease    ITMAX ,
the  spiral structure  of  spatial distribution
of SB is more evident ,
see  Figure~\ref{isosurf_spiral}~.

\begin{figure}
\begin{center}
\includegraphics[width=12cm,angle=0]{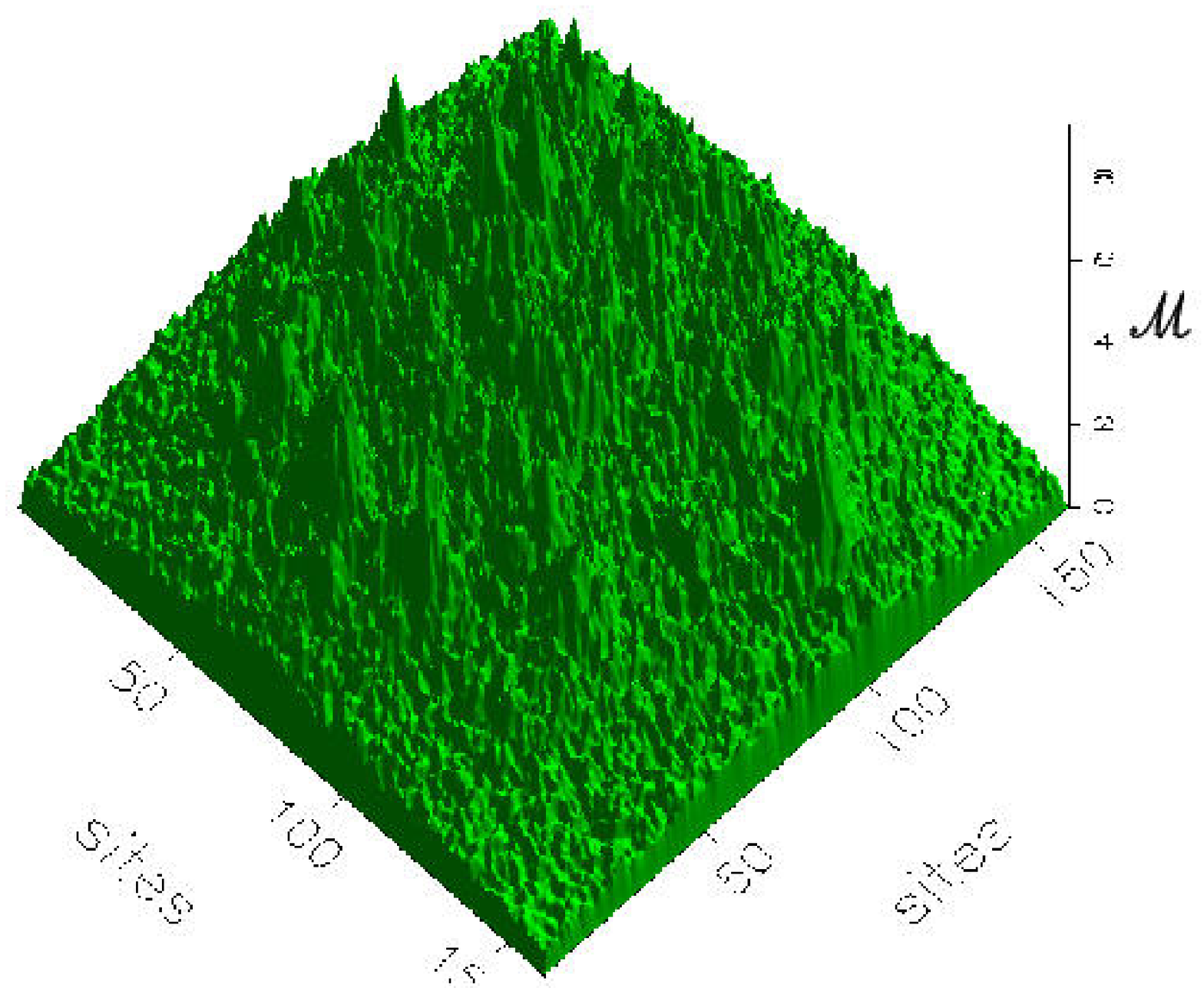}
\end {center}
\caption
{
 Concentration of CR
 at  the x-y (z=0)  Galactic plane. 
 Data as  in Figure~\ref{isosurf_plane}.
}
\label{xy_mountain}
    \end{figure}

\begin{figure}
\begin{center}
\includegraphics[width=12cm,angle=0]{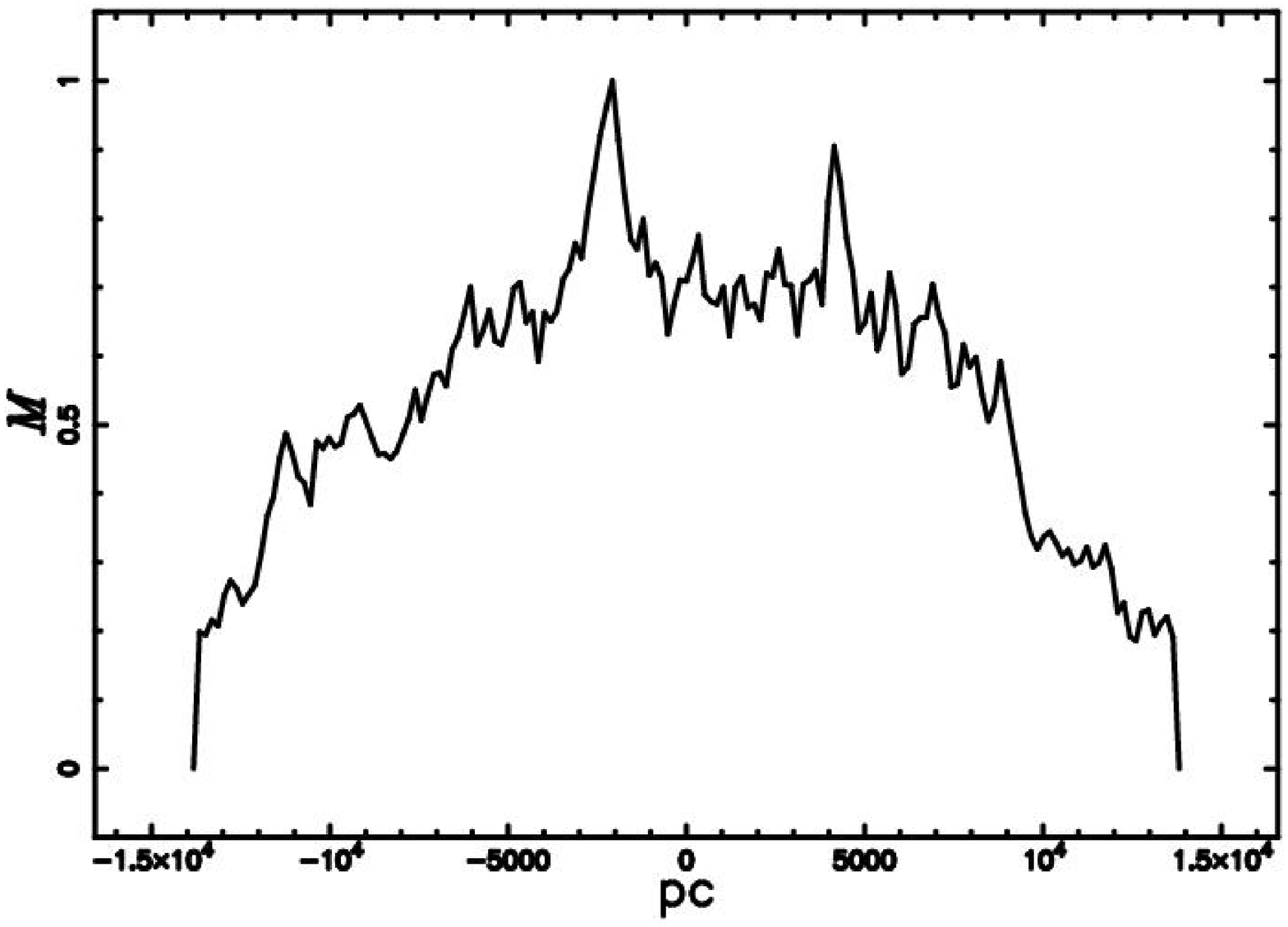}
\end {center}
\caption
{
 Cut along the Galactic plane , x-y,  of the concentration of CR
 crossing the center.
 Data as  in Figure~\ref{isosurf_plane}.
}
\label{linexy_plane}
    \end{figure}

The results can also be visualised through
 a 2D grid representing the concentration 
of  CR along the Galactic
plane x-y ,
 see  Figure~\ref{xy_mountain}, or 
1D cut , see  Figure~\ref{linexy_plane}~.

\begin{figure}
\begin{center}
\includegraphics[width=12cm,angle=0]{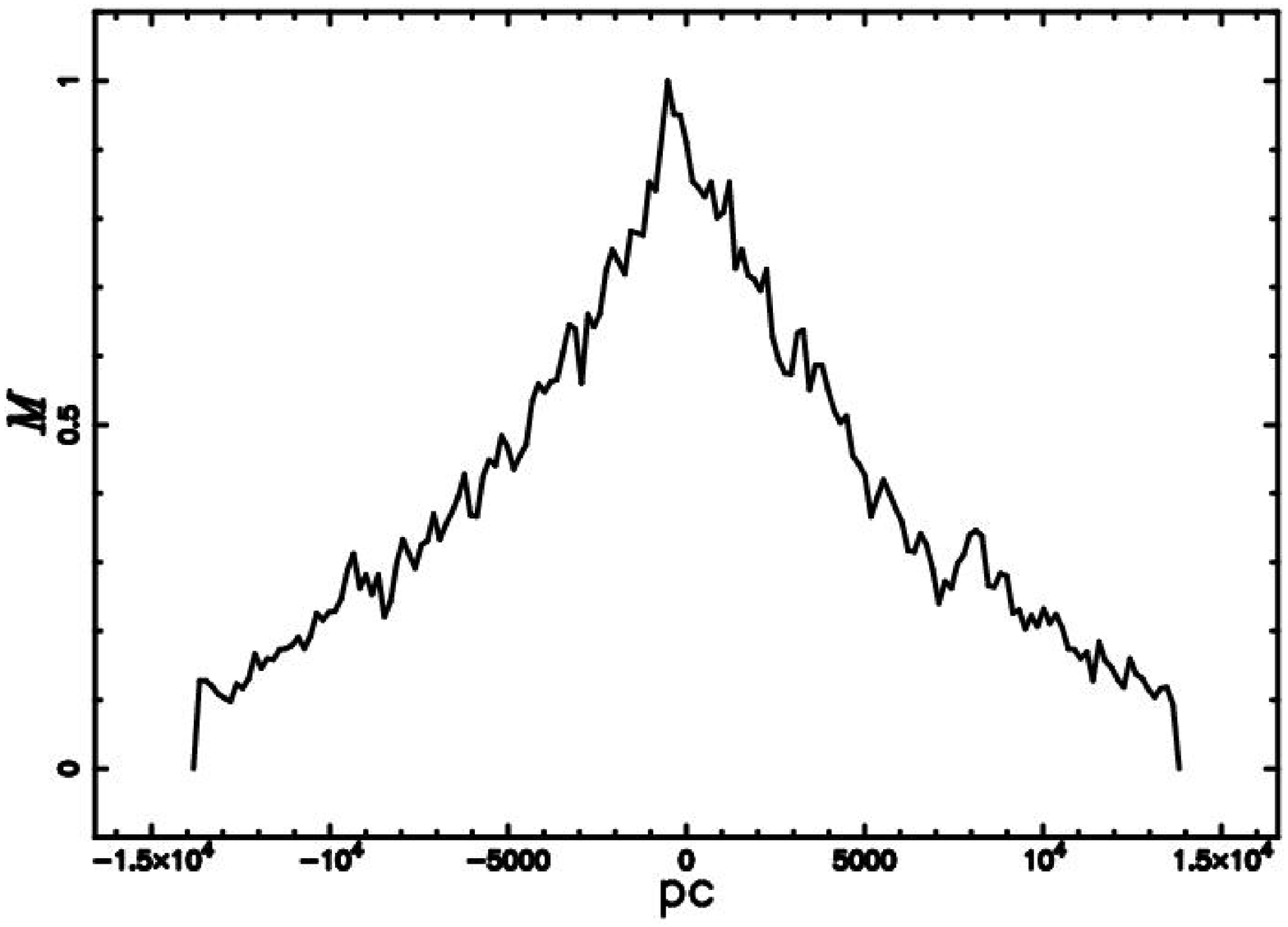}
\end {center}
\caption
{
 Cut of the concentration of CR
 along a direction perpendicular to the  Galactic plane
 and  crossing the center. 
 Data as  in Figure~\ref{isosurf_plane}.
}
\label{linexz_plane}
    \end{figure}

From this figure the contribution from
the SB belonging to the arms   is clear  ;
this  profile should be a flat line when the mathematical diffusion
is considered.
Another interesting case is the cut  along the Galactic height z ,
see  Figure~\ref{linexz_plane}.
from which the irregularities coming from the position of the
SB on the  arms are visible; this profile
should have a triangular  shape when the 1D  mathematical
diffusion from the Galactic plane is considered.

\subsection{The gamma emission from the Galactic plane}

The cosmic rays  with an energy range
$0.1~GeV < E < 4 10^5 GeV$ 
\cite{Hillas2005,Wolfendale2003} can produce gamma ray  emission
( $30 MeV < E < 30 GeV$)
from the interaction  with the target material
\cite{Wolfendale2003,Hillas2005,egret1997}.
The gamma ray emissivity will therefore be proportional
to the concentration of  CR.

\begin{figure}
\vspace {100pt}
\caption[]
{
Intensity of the gamma-rays     along the line of sight
when the galaxy is face on.
The parameters  of the diffusion are
Z=1                    ,
$t_D~=0.6~10^{7}$yr    ,
$c_{tr}=0.5 $          ,
$H_{-6}$=1             ,
$\lambda_1$=9.22  }    and 
( Bohm diffusion) $E= 8.5~10^{6}~GeV $  .
\includegraphics[width=12cm,angle=0]{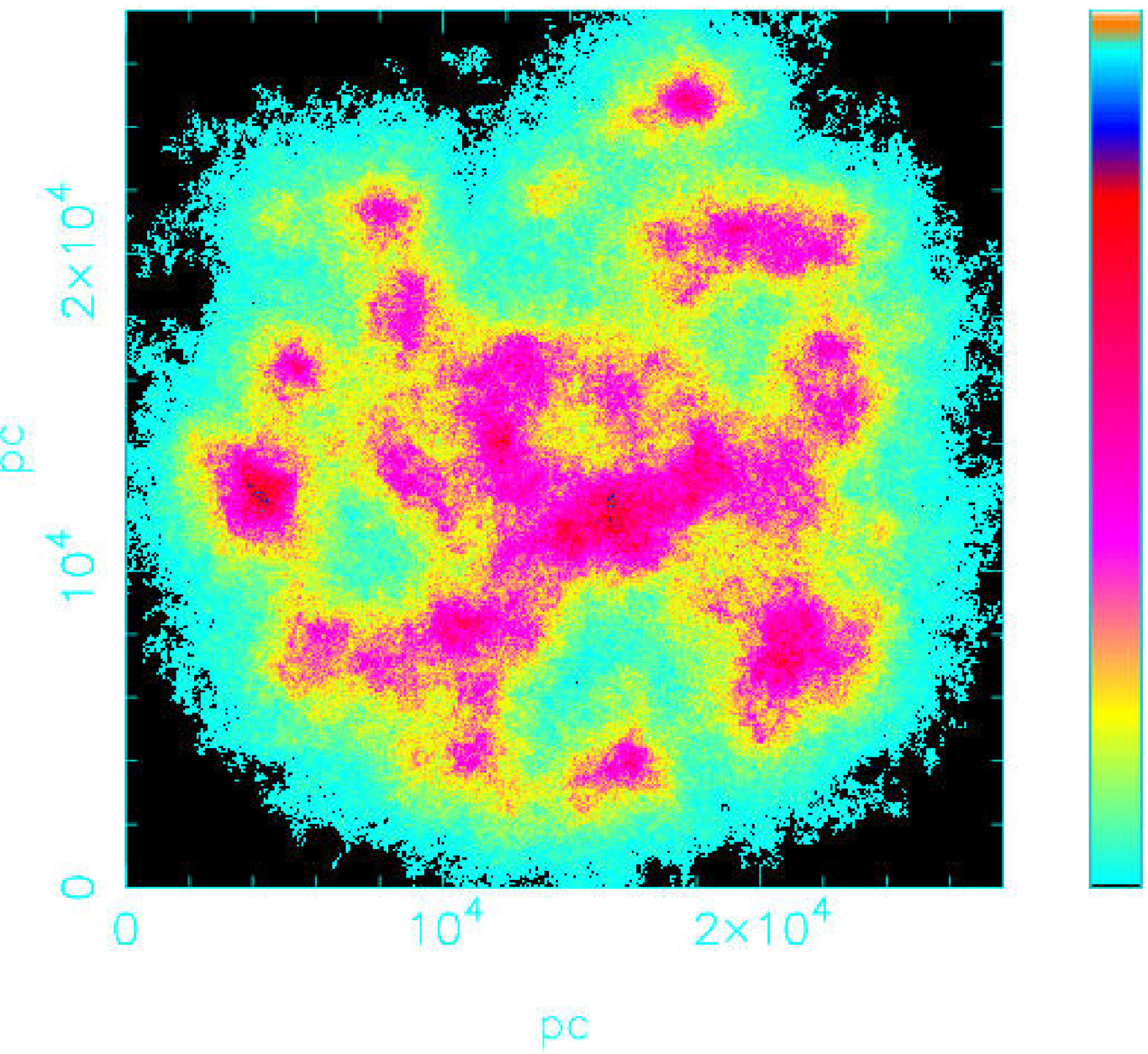}
\label{gammaz}%
    \end{figure}

\begin{figure}
\vspace {100pt}
\caption[]
{
Intensity of the gamma-rays     along the line of sight
in the Hammer-Aitof  projection.
The parameters  
are the same of Figure~\ref{gammaz}.
}
\includegraphics[width=12cm,angle=0]{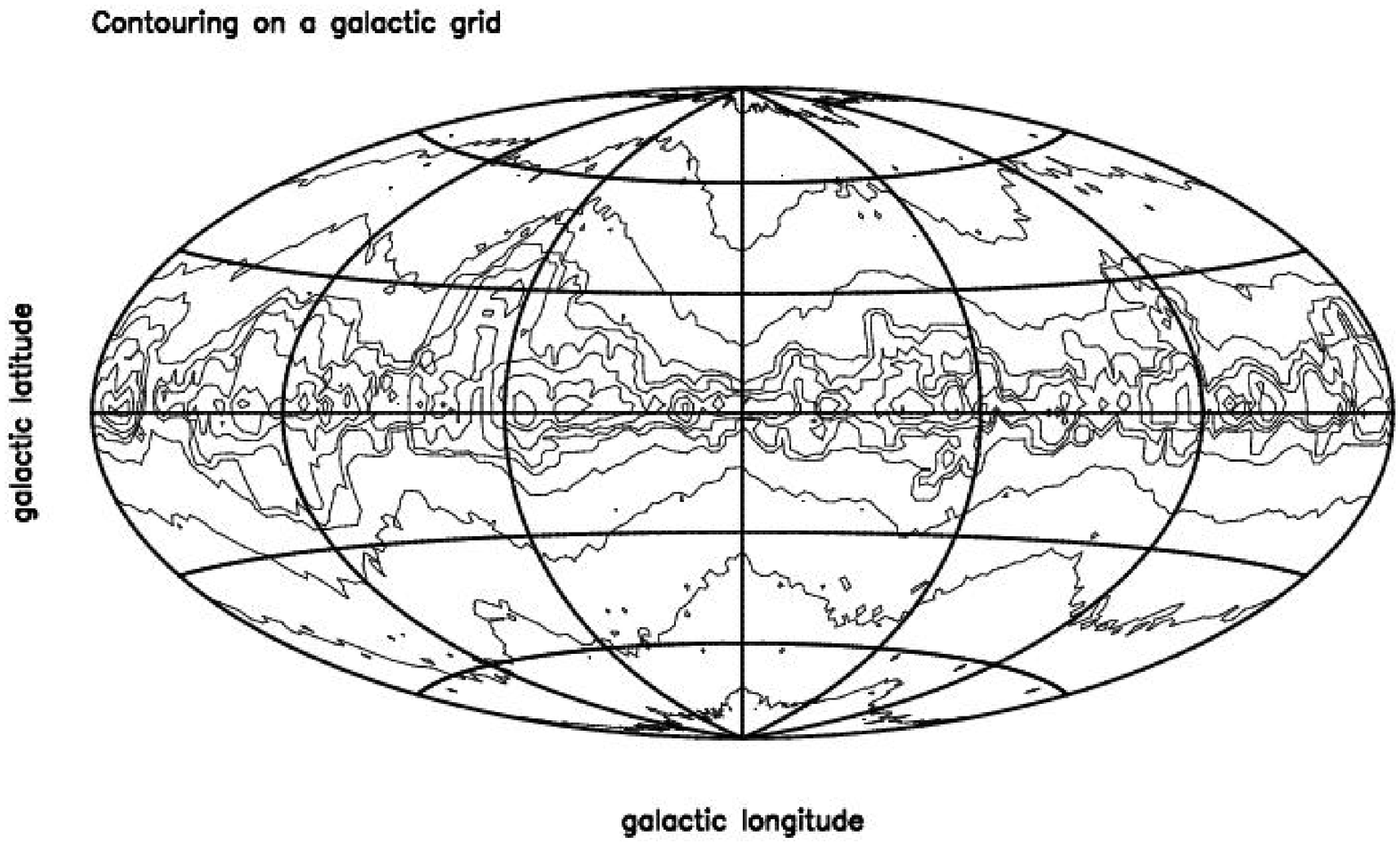}
\label{fighamm_log_nearest_gamma}%
    \end{figure}

A typical map of $\gamma$-emission from the Galactic plane is visualised
in  Figure~\ref{gammaz} , where the additive property
of the intensity of radiation along the line of sight is applied.
Using the   algorithm of the nearest IP~\cite{Zaninetti1988} it 
is  also possible 
to visualise the $\gamma$-emission
in the Hammer-Aitof  projection,see 
Figure~\ref{fighamm_log_nearest_gamma}.

\begin{figure}
\begin{center}
\includegraphics[width=12cm,angle=0]{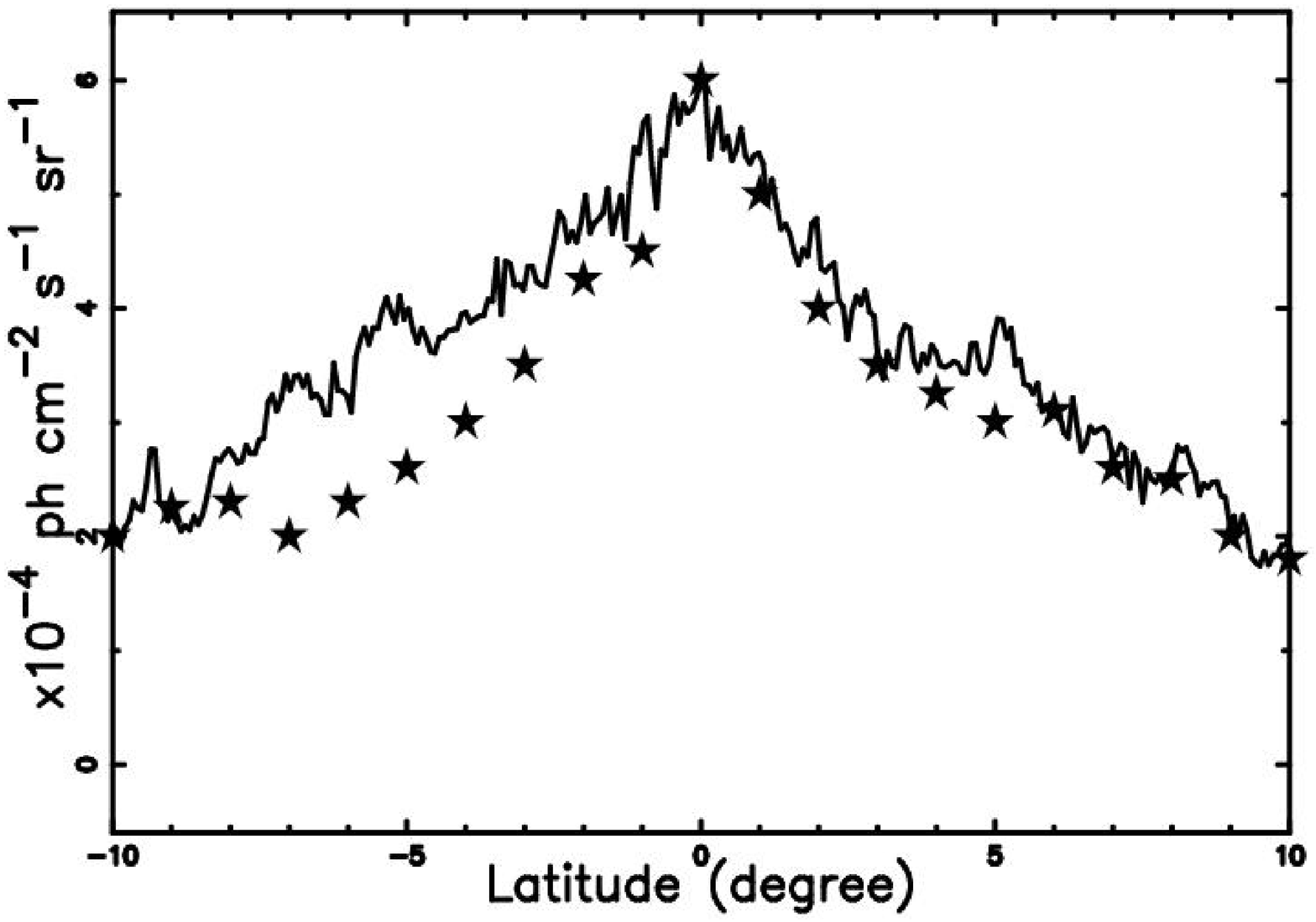}
\end {center}
\caption
{
Theoretical gamma-ray emission   as given by  
Monte Carlo simulation as function of the Galactic latitude
when the cut is crossing the center and the observer is at
700~pc,
data as  in Figure~\ref{gammaz}.
The stars represent the  observational cut. 
%
\label{latitude}}
    \end{figure}

The observations with EGRET~\cite{egret1997} 
provide  cuts in the intensity of radiation
that can be interpreted in the framework of the random walk.
We therefore report in  Figure~\ref{latitude}  the behaviour 
of the intensity of the gamma ray emission  proportional to 
concentration of CR when the distance from the Galactic 
plane is expressed in Galactic latitude;
in the figure  the stars represent  the  plot 
$(30^{\circ}>l>10^{\circ})$  of~\cite{egret1997}.

\subsection{Monte Carlo diffusion of CR from the Gould Belt  }

\label{gouldbelt}
\begin{figure}
\includegraphics[width=12cm,angle=0]{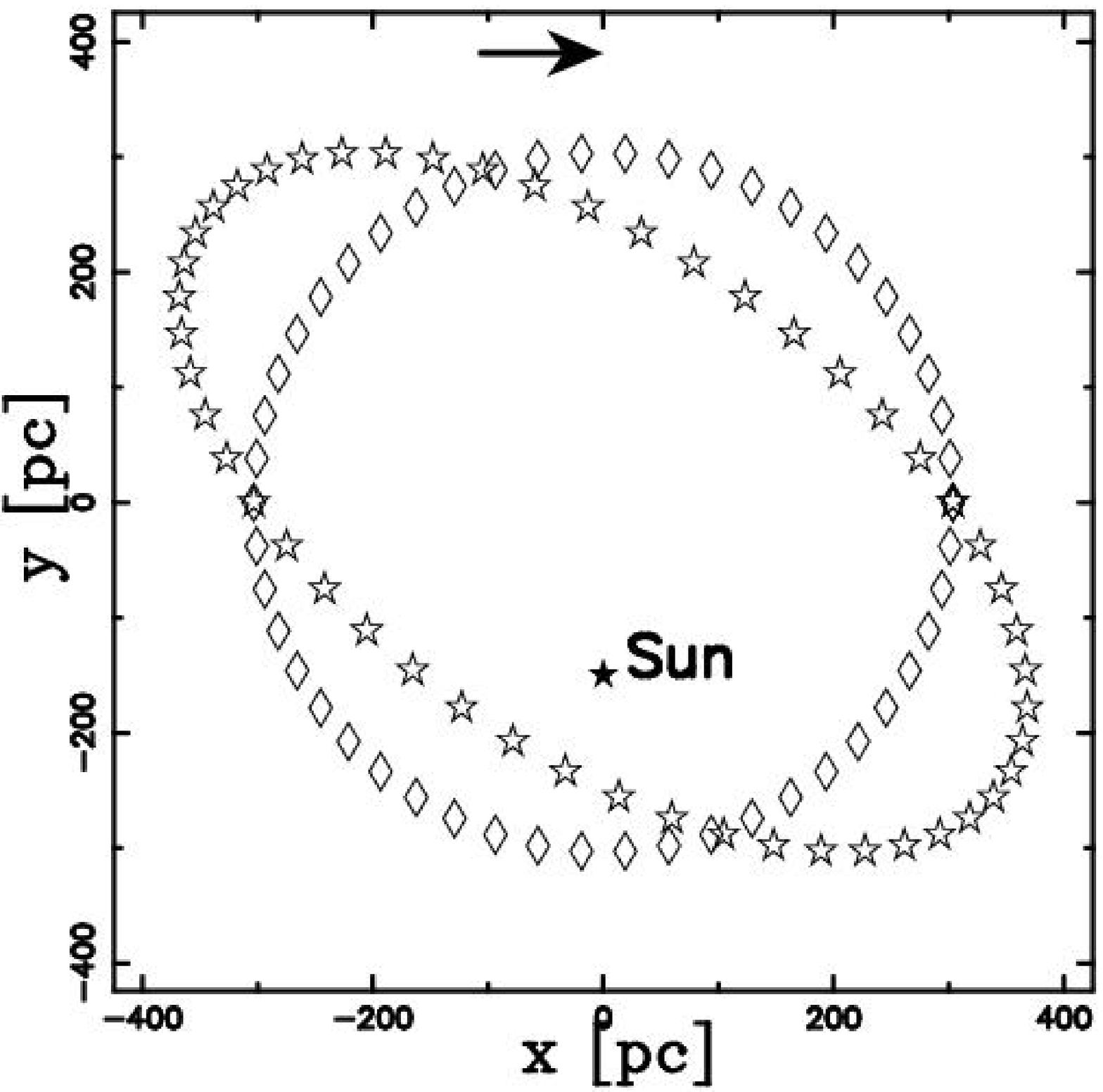}
\caption
{
Rhombi represent the circular  section,
the stars 
the rotation-distorted section and the big star the Sun.
The Galaxy direction of  rotation is also  shown.  
The parameters are
$t_\mathrm{\mathrm {age}}$=$2.6\cdot10^{7}~\mathrm{yr}$,  
$\Delta$~$t=0.001\cdot10^{7}~\mathrm{yr}$,
$t^{\mathrm{burst}}_7$= 0.015, $N^*$=  2000, $z_{\mathrm{OB}}$=0 pc,
and $E_{51}$=1.
}
\label{rotation_gould_sun}%
\end{figure}

\begin{figure}
\includegraphics[width=12cm,angle=0]{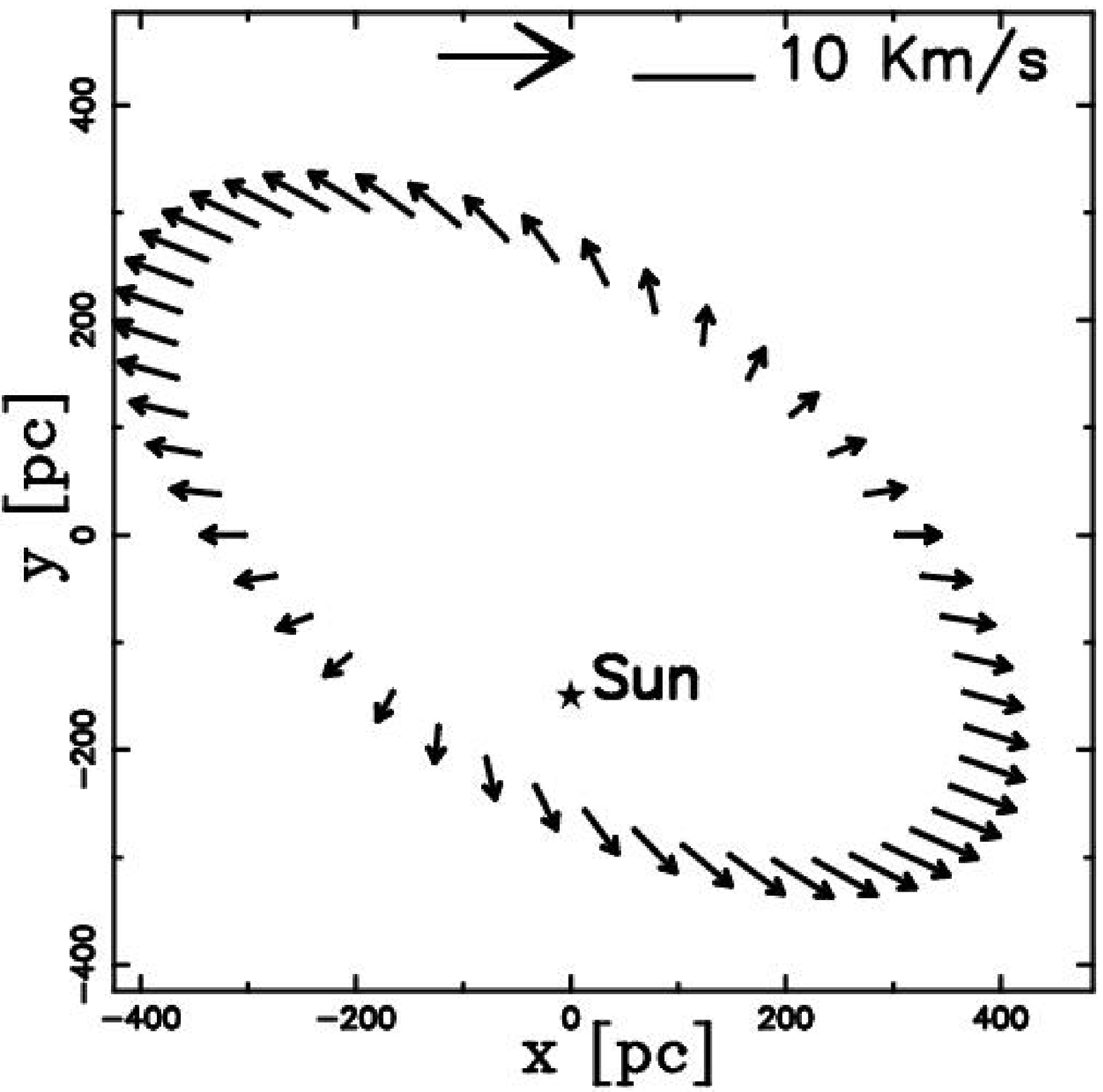}
\caption
{
The stars represents 
the rotation-distorted section of the Gould Belt  and the big star the Sun.
The velocity field   of  the expansion modified 
by the shear velocity  is mapped.
The Galaxy direction of  rotation is also  shown.  
}
\label{rotation_velocity}%
\end{figure}

  The physical 
parameters concerning  the Gould Belt    as  deduced in 
\cite{Perrot_2003},
  are reported in  Table~\ref{tab:gould}.
   \begin{table}
      \caption{Data of the super-shell associated with the Gould Belt}
         \label{tab:gould}
      \[
         \begin{array}{cc}
            \hline
            \noalign{\smallskip}
\mbox {Size (pc}^2)                    & 466 \cdot 746 \mbox{~at~b=0} \\
\mbox {Expansion~velocity~(km~s}^{-1})  & 17                \\
\mbox {Age~(10}^7\,{\mathrm{yr}})                  & 2.6              \\
\mbox {Total energy~\,(10}^{51}{\mathrm{erg}})      & 6               \\
            \noalign{\smallskip}
            \hline
         \end{array}
      \]
   \end{table}
The total energy  is  such as to produce  results  comparable
with the observations and  the  kinematic  age  is 
the same as in~\cite{Perrot_2003}.
In order to obtain 
$E_{\mathrm{tot}}=6.~10^{51}{\mathrm{erg}}$ 
with   $t^{\mathrm{burst}}=0.015~10^7~{\mathrm{yr}}$,
we have inserted 
$N^*$=~2000.
The time necessary to cross the Earth orbit , that lies
104 pc away from the Belt center, 
turns out 
to be  $0.078~10^7~yr$   that  means $2.52~10^7~yr$ from now.
The 2D  cut at {\it z}=0 of the superbubble  can be
visualised in~Figure~\ref{rotation_gould_sun}.
Our model gives a radial velocity at {\it z} =0  
$V_{\mathrm {theo}}$=3.67\,$\mbox {km~s}^{-1}$.
The influence of the Galactic  rotation on the direction and
modulus of the field  of originally radial velocity 
the transformation~(\ref{vshift}) is applied ,
see  Figure~\ref{rotation_velocity}.
A comparison should be done with Figure~5 
and Figure~9  in \cite{Perrot_2003}.

\begin{figure}
\vspace {100pt}
\caption[]
{
 Concentration of CR diffusing away from the Gould Belt
 at  the x-y (z=0) plane. 
$\lambda_1$=5.51, IP=10000 , ITMAX= 200 ,
$t_D~=~7204$~yr    and
$c_{tr}=0.5 $.
Assuming Z=1 ,$H_{-6}$=1 we have ( Bohm diffusion) $E_{15}$=5.08~.
 }
\includegraphics[width=12cm,angle=0]{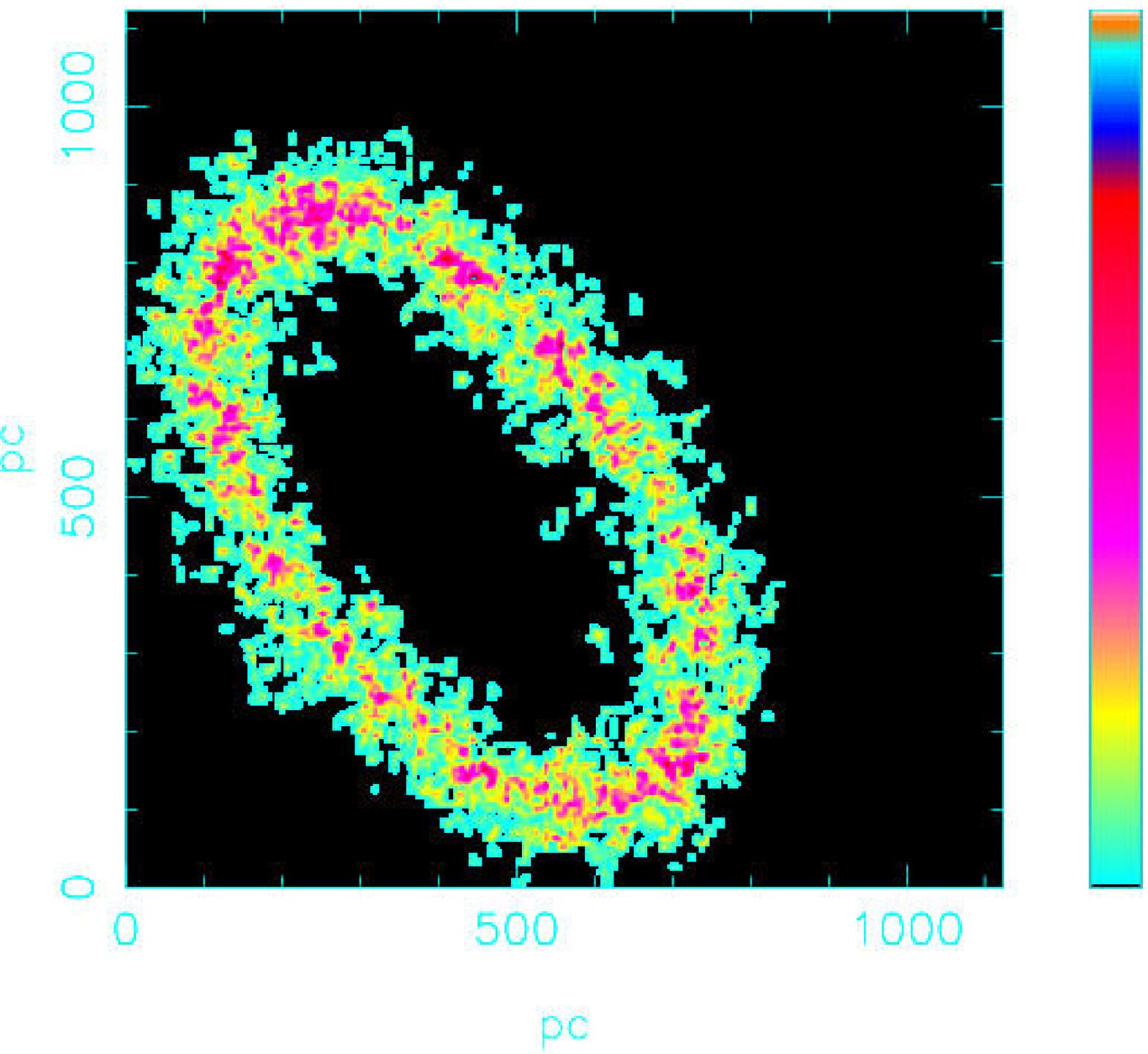}
\label{taglioxy}%
    \end{figure}

\begin{figure}
\vspace {100pt}
\caption[]
{
The same as  Figure~\ref{taglioxz} 
but in the
x-z (y=0) plane. 
}
\includegraphics[width=12cm,angle=0]{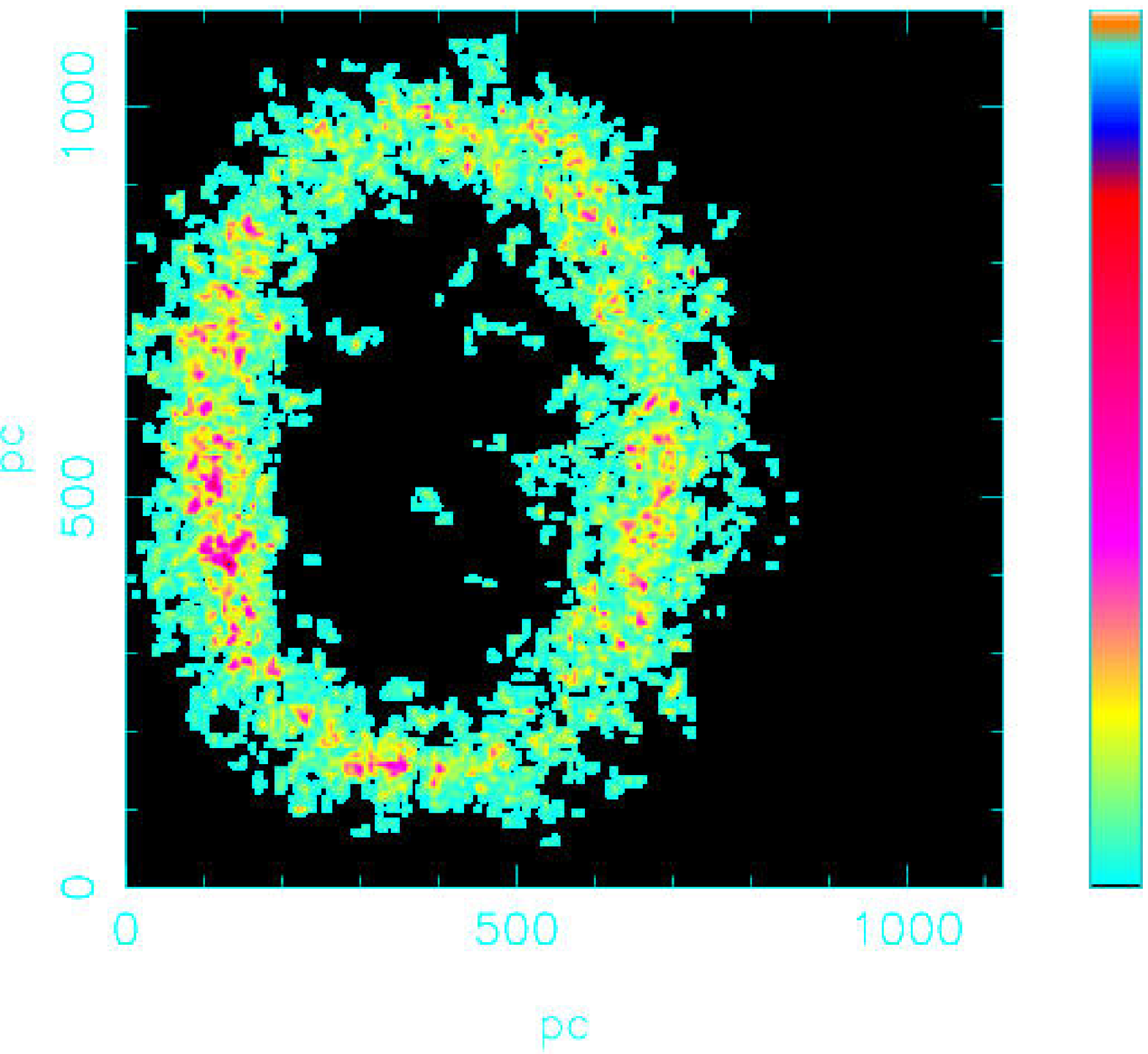}
\label{taglioxz}%
    \end{figure}

The results   of the streaming of CR from the complex 3D 
advancing surface   of the Gould Belt  can be visualized 
through a  2D grid representing the concentration 
of  CR along the Galactic plane   x-y (z=0) , see Figure~\ref{taglioxy},
or through  a  perpendicular plane x-z (y=0), see Figure~\ref{taglioxz}.

The variation of CR due to the presence of the Gould Belt are also 
analyzed in Section 3.2 in \cite{Schlickeiser2005}.

\section{Conclusions}

We have studied the propagation  of CR with the aid of
mathematical diffusion in 1D, 2D and 3D. Due to the nature of the
problem , diffusion from the Galactic plane , the  first results can be
obtained in 1D. The 1D diffusion has also been analysed in the
case of a variable diffusion coefficient: which  in our astrophysical
case consists in a magnetic field function of the altitude z from
the Galactic plane. These analytical results  allow to
test the Monte Carlo diffusion. The physical explanation is the
diffusion through  the Larmor radius   which  means that the diffusion
coefficient depends on  energy. The calculation of the transit
times of CR explains   the absence of  the isotope  $~^{10}Be$. 
Further on the theoretical  and Monte Carlo set up  of the Levy
flights allows to increase the  applications of the random walk.
For example from an observed  profile of CR function of distance
from the Galactic plane is possible to evaluate the concavity:
the 1D Levy flights predict profiles concave up and the 1D regular 
random walk predicts no concavity. 
The observations with  EGRET in the gamma range
, see Figure~3 in ~\cite{egret1997} , 
provide  profiles that are concave up and therefore 
the Levy flights with the shape regulating 
  parameter $\alpha$ can be an explanation
 of the observed profiles, see Figure~\ref{cut1d}.

The
CR diffusion in a super-bubble environment is then performed by 
evaluating the maximum energies that can be extracted
($\approx~10^{17}eV) $ and  performing a Monte Carlo
simulation of the propagation of CR from SB disposed on a
spiral-arm network. According to the numerical results,
 the
concentration of CR depends on the distance from the nearest SB
and from the chosen energy. 

This fact favours the local origin of CR \cite{Wolfendale2005} that
comes out from an   
analysis of the results of KASCADE data \cite{Hoerandel2004}.
From an  astrophysical point of view, local origin means 
to model the diffusion from the nearest SNR or SB.
Like a practical  example we now explore the case of the Gould Belt,
the nearest SB , which  has an average distance of 300~$\approx$~pc 
from us. 
The concentration of CR at the source of this SB is ,
according to formula~(\ref{approximate3D}):
\begin {equation}
C_{{m}} \approx C(300) \frac {300~pc}{\rho_Z}
 \approx 277 \times C(300) \frac   {H_{-6} Z  }{E_{15}} 
\quad ,
\end {equation}
where $C(300)$ is the measured concentration on Earth 
at the given energy.
Is interesting to point out that the crossing time  
(25~My~ago)
of the expanding Gould
Belt with the Earth  with  a  consequent increases
in the concentration of CR  can explain one event   
of the fossil diversity cycle  that
has a periodicity of 62~My  \cite{Melott_2006}~.
The gamma-ray emissivity is different
in the various regions of the galaxy \cite{Digel2001} , and in
order to explain this fact the theoretical map of gamma-ray from
the Galactic plane has been produced.
CR with energies greater 
that $10^{18}eV$ , Ultra-High Energy (UHE) , 
are thought to be of extra-galactic origin 
(\cite{Sigl_2000,Dolag_2005,DeMarco_2006,Gorbunov_2006,Tkachev_2003,Westerhoff_2006})
and therefore the diffusion  from extra-galactic sources
through the intergalactic medium  
 is demanded to a future
investigation.
\section*{Acknowledgments}
I thank M. Ferraro  for many fruitful discussions.

\end{document}